\pdfoutput=1
\documentclass[amssymb,amsmath,prd,nofootinbib,twocolumn,superscriptaddress]{revtex4-1}
\usepackage{fontawesome5}
\usepackage{float}
\usepackage{lipsum}
\usepackage{bbm}
\usepackage{appendix}
\usepackage{diagbox}
\usepackage{booktabs}
\usepackage{physics}
\usepackage{amsmath}
\usepackage{amssymb}
\usepackage{amsfonts}
\usepackage{mathrsfs}
\usepackage{tcolorbox}
\usepackage[nolist,nohyperlinks]{acronym}
\usepackage[caption=false]{subfig}
\usepackage{url}
\usepackage{multirow}
\usepackage{booktabs}
\usepackage{xcolor}
\definecolor{linkcolor}{HTML}{4A70A9}
\usepackage[%
  colorlinks = true,
  urlcolor= linkcolor,
  linkcolor= linkcolor,
  citecolor= linkcolor
]{hyperref}
\usepackage{orcidlink}
\usepackage{float}
\usepackage{graphicx}
\usepackage{capt-of}
\usepackage{tikz-cd}
\usepackage{tikz}
\usetikzlibrary{shapes,arrows}
\usetikzlibrary{shapes.geometric, arrows}
\usetikzlibrary{graphs}
\usetikzlibrary{cd}
\usepackage{dcolumn}
\usepackage{bm}
\usepackage{booktabs}
\newcommand{\D}{\mathrm{d}}
\newcommand{\p}{\partial}
\newcommand{\OGW}{\Omega_\mathrm{GW}}

\begin{document}

\title{\texttt{StochAXtic}: a JAX-Accelerated, End-to-End Differentiable Framework for Gravitational-Wave Backgrounds from Compact Binary Populations}

\author{Haowen Zhong\texorpdfstring{\,}{}\orcidlink{0000-0001-8324-5158}}
\email{hzhong23@jh.edu}
\affiliation{William H. Miller III Department of Physics and Astronomy, Johns Hopkins University, Baltimore, Maryland 21218, USA}

\author{Ethan Vislosky\texorpdfstring{\,}{}\orcidlink{0009-0007-3950-6411}}
\email{evislos1@jh.edu}
\affiliation{William H. Miller III Department of Physics and Astronomy, Johns Hopkins University, Baltimore, Maryland 21218, USA}

\author{Emanuele Berti\texorpdfstring{\,}{}\orcidlink{0000-0003-0751-5130}}
\email{berti@jhu.edu}
\affiliation{William H. Miller III Department of Physics and Astronomy, Johns Hopkins University, Baltimore, Maryland 21218, USA}
\date{\today}
\begin{abstract}
We present \texttt{StochAXtic}, a package for computing the dimensionless energy density spectrum $\OGW(f)$ of the stochastic gravitational-wave background generated by compact binary coalescences. Rather than sampling binary populations via Monte Carlo, \texttt{StochAXtic} exploits the \emph{scale invariance} of the waveforms: at fixed mass ratio, spins and tidal deformabilities, the waveform depends on the total mass only through the dimensionless combination $Mf$.  This allows all computationally expensive waveform evaluations using \texttt{LALSuite} to be performed once and stored in a population-independent lookup table. The subsequent integrals over mass and redshift can then be recast as convolutions and evaluated efficiently using fast Fourier transforms. We further employ truncated singular value decomposition to compress the lookup table, reducing both computational cost and memory requirements. 
Once the table has been computed and cached, \texttt{StochAXtic} evaluates $\OGW(f)$ in a few milliseconds, achieving a relative numerical error below $10^{-3}$ at frequencies up to $1$ kHz. Its results agree with those of \texttt{popstock} within the latter's statistical uncertainties.
Implemented entirely in \texttt{JAX}, \texttt{StochAXtic} also provides gradients of $\OGW(f)$ with respect to population hyperparameters, enabling its direct integration with gradient-based inference methods including Hamiltonian Monte Carlo.
\end{abstract}
\maketitle
\begin{acronym}
    \acro{GW}{gravitational-wave}
    \acro{ORF}{overlap reduction function}
    \acro{PSD}{power spectral density}
    \acro{GR}{general relativity}
    \acro{CBC}{compact binary coalescence}
    \acro{BH}{black hole}
    \acro{NS}{neutron star}
    \acro{BBH}{binary black hole}
    \acro{BNS}{binary neutron star}
    \acro{NSBH}{neutron star-black hole}
    \acro{LVK}{LIGO-Virgo-KAGRA}
    \acro{PE}{parameter estimation}
    \acro{SGWB}{stochastic gravitational wave background}
    \acro{CGWB}{cosmological gravitational wave background}
    \acro{CDF}{cumulative distribution function}
    \acro{CGF}{cumulant generating function}
    \acro{MGF}{moment generating function}
    \acro{MC}{Monte Carlo}
    \acro{CE}{Cosmic Explorer}
    \acro{ET}{Einstein Telescope}
    \acro{XG}{next-generation}
    \acro{SNR}{signal-to-noise ratio}
    \acro{CLT}{central limit theorem}
    \acro{HDI}{highest density interval}
    \acro{PN}{post-Newtonian}
    \acro{PTA}{pulsar timing array}
    \acro{FFT}{Fast Fourier Transform}
    \acro{SVD}{singular value decomposition}
    \acro{EOS}{equation of state}
    \acro{QMC}{quasi Monte Carlo}
    \acro{MCMC}{Markov Chain Monte Carlo}
    \acro{NUTS}{No-U-Turn Sampler}
\end{acronym}
\section{Introduction}
A  \ac{SGWB} is the superposition of a large number of incoherent \acp{GW} from sources throughout the entire cosmic history~\cite{Christensen:2018iqi, Romano:2016dpx, Renzini:2022alw}. 
An \ac{SGWB} can be astrophysical~\cite{Regimbau:2011rp,Regimbau:2022mdu}, produced for instance by \ac{CBC} events~\cite{Zhu:2011bd,Rosado:2011kv}; or cosmological~\cite{Caprini:2018mtu}, produced by first-order phase transitions~\cite{Caprini:2015zlo,Caprini:2019egz}, cosmic strings~\cite{Damour:2004kw, Siemens:2006yp}, domain walls~\cite{Saikawa:2017hiv}, and inflation~\cite{Starobinsky:1979ty, Guzzetti:2016mkm}.
Consequently, an \ac{SGWB} provides a natural probe of astrophysics, cosmology and particle physics. The astrophysical background carries information on the merger history of compact binaries far beyond the horizon of individual detections~\cite{Callister:2016ewt,Callister:2020arv}, while observations of a cosmological background would probe the early Universe and physics beyond the Standard Model at energy scales far above the reach of colliders~\cite{Caprini:2018mtu}. 

The standard search for an \ac{SGWB} cross-correlates the data from detector pair(s) ~\cite{Allen:1997ad, Romano:2016dpx, Zhong:2026owg}, where the expected value of the cross-correlation estimator is equal by construction to the dimensionless energy density spectrum $\OGW(f)$. The latest isotropic search of the \ac{LVK} Collaboration, using data collected through April 2025, reports no detection and therefore places an upper limit of $\OGW(25\,\rm{Hz})\leqslant 2.0\times 10^{-9}$ on the stochastic background from inspiraling binaries~\cite{LIGOScientific:2026mjf}. This limit is only about three times the predicted \ac{CBC} background, $\Omega_{\rm CBC}(25\,\rm{Hz})=6.3^{+5.0}_{-2.2}\times 10^{-10}$~\cite{LIGOScientific:2026mjf}, suggesting that the first hint of an astrophysical background may be within reach during the upcoming fifth observing run (O5) of the \ac{LVK} Collaboration, when the LIGO detectors are expected to operate at their upgraded A$+$ design sensitivity. 
Given this non-detection, in Ref.~\cite{KAGRA:2025swt} the \ac{LVK} Collaboration considered also various \ac{CGWB} models and placed constraints on the parameters of those models.

\begin{figure*}[!htbp]
    \centering
    \includegraphics[width=0.95\linewidth]{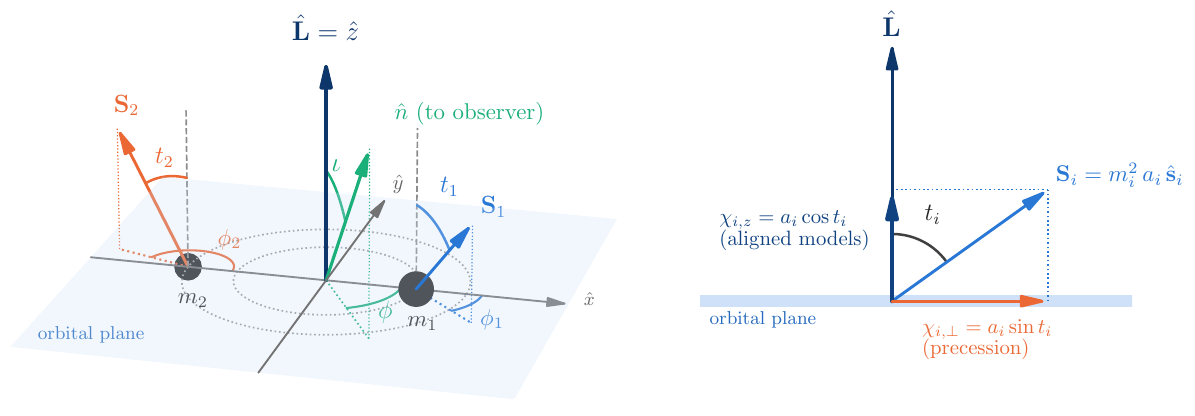}
    \caption{Sketch of the geometry of a binary system. Left: two compact objects (\ac{BH} or \ac{NS}) with masses $m_1$ and $m_2$, spins $\bm S_1$ and $\bm S_2$. We define the $\hat z$ direction as the direction of the orbital angular momentum $\hat {\bm L}$. The angle between the individual spins and $\hat z$ is denoted by $t_i$, $i\in\{1,2\}$. The $\hat x$-axis is along the line joining two objects (pointing from $m_2$ to $m_1$). The $\hat y$ axis is chosen such that $\hat y=\hat z\times \hat x$; $\hat n$ denotes the unit vector pointing from the binary to the observer, whose tilt and azimuthal angles are $\iota$ and $\phi$. The azimuthal angles of $\bm S_1$ and $\bm S_2$ are denoted by $\phi_1$ and $\phi_2$, respectively. Right: one can decompose both $\bm S_1$ and $\bm S_2$ into components parallel or perpendicular to the $\hat{\bm L}$ direction.}
    \label{fig:binary_angles}
\end{figure*}

Detecting an \ac{SGWB} is a major science goal of current \ac{GW} detectors. For next-generation ground-based detectors such as \ac{CE}~\cite{CEHorizon} and \ac{ET}~\cite{Maggiore2020ET, Abac:2025saz}, however, the \ac{CBC} background will be detected with a cross-correlation \ac{SNR} of order $10^3$. These detectors are also expected to resolve almost all \ac{BBH} mergers and about half of the \ac{BNS} mergers whose \ac{GW} signals reach Earth during the observing period~\cite{CEHorizon, Abac:2025saz}. The focus will then shift from detecting the \ac{CBC} background to searching for a cosmological \ac{SGWB} beneath it, and the \ac{CBC} signals will become a bright foreground that contaminates that search~\cite{Regimbau:2016ike,Zhou:2022otw,Zhou:2022nmt,Song:2024pnk,Zhong:2022ylh,Zhong:2024dss,Zhong:2025qno,Biscoveanu:2020gds,Pan:2023naq,Li:2024iua}. To address this challenge, \citet{Zhong:2025qno} proposed to first notch out the resolvable events~\cite{Zhong:2022ylh,Zhong:2024dss}, and then infer the \ac{CBC} population and the cosmological \ac{SGWB} jointly, following the framework proposed in~\citet{Callister:2020arv}. Later,  \citet{Cousins:2025bas} proposed a so-called ``stochastic siren'' search, which adopts a similar framework to simultaneously constrain the \ac{CBC} population and the Hubble constant. We emphasize that both analyses require $\OGW(f)$ of the \ac{CBC} background to be recomputed for every population drawn by the sampler in real time. 

In these studies, $\OGW(f)$ is typically approximated by the Newtonian inspiral spectrum, $\OGW(f)\propto f^{2/3}$. This approximation neglects the merger and ringdown phases and cannot capture the effects of spins, precession, and tidal interactions, nor accommodate different waveform models. Alternatively, $\OGW(f)$ can be computed using \ac{MC} methods. The \texttt{popstock}  package~\cite{Renzini:2024pxt} constructs a bank of \ac{CBC} events drawn from a fiducial population and uses importance sampling to reweight the events when evaluating $\OGW(f)$ for target populations. The statistical uncertainty of this approach depends on the size of the event bank and grows as the target population deviates from the fiducial one. Moreover, incorporating additional physical effects requires a larger event bank, increasing both the memory requirement and the computational cost of each evaluation. 

In this work we present \texttt{StochAXtic}, a new package that efficiently computes $\OGW(f)$ and overcomes these limitations. The numerical implementation is described in Sec.~\ref{sec:numerical_implementations}. We start from the simplest case, in which only the mass and redshift distributions of the \ac{CBC} events are modeled (Sec.~\ref{sec:m_z}). We then add spins (Sec.~\ref{sec:m_z_s}; aligned spins in Sec.~\ref{sec:m_z_s_aligned}, and precessing spins in Sec.~\ref{sec:m_z_s_precession}) and tidal effects (Sec.~\ref{sec:m_z_s_t}). 
In Sec.~\ref{sec:convergence} we present convergence tests, and in Sec.~\ref{sec:sim} we carry out further numerical tests. In Sec.~\ref{sec:discussion_conclusion} we discuss our results and conclusions.

Technical details are collected in Appendices~\ref{app:average_over_n}, \ref{app:SVD}, \ref{app:preconditioning}, \ref{app:QMC} and \ref{app:n_w}. Appendix~\ref{app:supported} lists the population and waveform models supported by \texttt{StochAXtic}, together with their default parameters.
\section{Numerical Implementation}\label{sec:numerical_implementations}
In this section we discuss the numerical implementation of \texttt{StochAXtic}. 
Our aim is to compute the dimensionless energy density spectrum from various \ac{CBC} populations,
\begin{equation}
    \OGW(f)\equiv\frac{f}{\rho_c}\frac{\D\rho_{\rm GW}(f)}{\D f},
\end{equation}
where $\D\rho_{\rm GW}(f)$ is the GW energy density in the frequency band $(f\,,f+\D\!f)$, and $\rho_c\equiv 3H_0^2c^2/(8\pi G)$ is the critical energy density today corresponding to a spatially flat Universe.

In the first subsection, we consider the simplest case where only mass and redshift distributions of \ac{CBC} populations are taken into account; in the second subsection, we consider non-zero spins; in the third subsection, we incorporate tidal effects into our framework. To simplify the derivation, we use geometrical units ($G=c=1$), switching to SI units if needed. 

Before delving into the details, in Fig.~\ref{fig:binary_angles} we present a simple sketch of the geometry of a binary system. The two compact objects are labeled by their source-frame masses $m_1$ and $m_2$, with $m_1\geqslant m_2$. We define the line joining $m_2$ and $m_1$ as the $x$-axis, with the positive $\hat x$ direction pointing from $m_2$ to $m_1$. We choose the $z$-axis to be aligned with the orbital angular momentum $\bm L$. Both objects can have non-zero and arbitrarily oriented spins $\bm S_1$ and $\bm S_2$. We define the tilt angles between the individual spin vectors and the orbital angular momentum as $t_1$ and $t_2$. The azimuthal angles of the two spins are then denoted by $\phi_1$ and $\phi_2$, with $\phi_{12}\equiv \phi_1-\phi_2$;
$\hat n\equiv(\iota,\phi)$ denotes the unit vector pointing from the binary to the observer. A change of the coalescence phase $\phi_c$ is then equivalent to a rotation of this frame about $\bm L$, i.e., to a change of $\phi$, so $\phi_c$ is not an independent parameter. 
The right panel of Fig.~\ref{fig:binary_angles} shows the decomposition of $\bm S_i$ into parallel and perpendicular components with respect to $\bm L$, where $\chi_{i,z}=a_i\cos t_i$ denotes the dimensionless spin of the compact object $i$, which is aligned with $\bm L$, whereas $\chi_{i,\perp}=a_i\sin t_i$ denotes the component perpendicular to $\bm L$, which results in precession of the orbital angular momentum $\bm L$ about the total angular momentum $\bm J\equiv \bm L+\bm S_1+\bm S_2$.
We note that all these angles are defined at a reference frequency $f_{\mathrm{ref}}$. For the non-precessing case, choosing a different $f_{\rm ref}$ does not make any difference, whereas in the precessing case, defining the frequency at a fixed physical frequency can make the calculation complicated. We will discuss the impact of precession in  Sec.~\ref{sec:m_z_s_precession}.

\subsection{Case I: Mass and redshift distributions}\label{sec:m_z}
To begin with, we neglect spins and tidal deformabilities of the compact objects. A binary is then described by $\bm\theta=\{m_1,m_2,d_L(z),\hat n,t_c,\psi,\alpha,\delta;\,\bm\chi_1=\bm\chi_2=\bm 0,\Lambda_1=\Lambda_2=0\}$, where $m_1, m_2,\hat n,\bm\chi_1,\bm\chi_2$ have been defined above;$d_L(z)$ is the luminosity distance of the event, which is a function of redshift $z$ for a given cosmological model (throughout this work, we adopt the \texttt{Planck18} cosmology~\cite{Planck:2018vyg} implemented in \texttt{Astropy}~\cite{Astropy:2013muo,Astropy:2018wqo,Astropy:2022ucr});
$t_c$ is the coalescence time, $\psi$ the polarization angle, $\alpha$ and $\delta$ the right ascension and declination, and $\Lambda_1$, $\Lambda_2$ are the tidal deformabilities of the two objects. The coalescence phase does not appear separately, as it is already absorbed into the definition of the azimuth $\phi$ of $\hat n$.

We note that $t_c,\psi,\alpha$, and $\delta$ leave the energy density spectrum unchanged. Therefore, $\OGW(f)$ only requires an average over the population distributions of the masses and redshift, and orientation $\hat n$~\cite{Phinney:2001di,Belgacem:2024ohp,Renzini:2024pxt, Zhong:2026jzl}:
\begin{equation}
    \OGW(f)=\frac{4\pi^2 f^3}{3H_0^2}\int_{z_\mathrm{min}}^{z_\mathrm{max}}\D z\,\frac{R(z)}{1+z}\frac{\D V_c}{\D z}\left\langle \mathcal{I}(f;\bm\theta)\right\rangle_{m_1,\,m_2,\,\hat n},
    \label{eq:integral_stage0}
\end{equation}
where $\D V_c/\D z$ denotes the comoving volume element per unit redshift for the assumed cosmological model, and $\mathcal{I}(f;\bm\theta)\equiv |\tilde h_+(f;\bm\theta)|^2+|\tilde h_\times(f;\bm\theta)|^2$ the \ac{GW} intensity of a source with parameters $\bm\theta$, viewed along the source-frame direction $\hat n$. The brackets
$\langle \cdots\rangle_{X}$ denote the average over the distribution of $X$; explicitly,

\begin{equation}
    \langle \cdots\rangle_X\equiv\int\D X~p(X)[\cdots],
\end{equation}
where $X$ can be multi-dimensional, i.e., a random vector, and $p(X)$ is the probability density function (PDF) over $X$.

We stress that there is \emph{no intrinsic scale} in vacuum Einstein gravity~\cite{Lehner:2014asa}, so a waveform for one total mass $M$ maps trivially into any other: $M$ simultaneously sets both the time scale and the amplitude of the emitted waveform~\cite{Aylott:2009ya}. In the case when only the mass and redshift distributions (no spins or tides) of \ac{CBC} populations are considered, \textit{scale invariance}, together with the inverse-distance falloff of the strain in the radiation zone, allows us to write the \ac{GW} strain emitted by a \ac{CBC} with total mass $M$, mass ratio $q$, and distance $D$ as~\cite{Aylott:2009ya}
\begin{equation}
    h(t;M,q,\hat n, D)=\frac{M}{D}H\left(\frac{t}{M};q,\hat n\right),
    \label{eq:pillar}
\end{equation}
where $H(\cdot)$ is a function of $t/M$, mass ratio $q$, and orientation $\hat n$ (this is no longer the case when additional scales enter the problem). 
At fixed mass ratio and orientation, $M/D$ sets the amplitude, while time enters as $t/M$. Given this relation, one can perform a Fourier transform to get
\begin{equation}
    \tilde h(f;M,q,\hat n,D)= \frac{M^2}{D}\tilde H(Mf;q,\hat n).
    \label{eq:scale_invariance}
\end{equation}
In SI units, one has $Mf\to GMf/c^3$. Here, we define
\begin{align}
    x\equiv\frac{GMf}{c^3}=\frac{M}{M_\odot}f\frac{GM_\odot}{c^3}\equiv\frac{M}{M_\odot}fT_\odot,
\end{align}
with $T_\odot\equiv GM_\odot/c^3\approx4.925\,\mathrm{\mu s}$.
Inserting Eq.~\eqref{eq:scale_invariance} back into the definition of $\mathcal I(f;\bm\theta)$ then yields
\begin{align}
    \mathcal{I}(f;\bm\theta)=\frac{M^4}{D^2}\left(|\tilde H_+(x;q,\hat n)|^2+|\tilde H_\times(x;q,\hat n)|^2\right).
\end{align}

To restore $G$ and $c$, we have
\begin{align}
    \frac{M^4}{D^2}\to \frac{G^4M^4}{c^{10}D^2}=\frac{G^4M_\odot^4}{c^{10}(1\,\mathrm{Mpc})^2}\frac{(M/M_\odot)^4}{(D/1\,\mathrm{Mpc})^2}.
\end{align}

\begin{widetext}
Now, the expression of $\mathcal{I}(f;\bm\theta)$ reads
\begin{equation}
    \mathcal{I}(f;\bm\theta)=\left(\frac{c^2T_\odot^4}{ 1\,\mathrm{Mpc}^2}\right)\frac{(M/M_\odot)^4}{(D/1\,\mathrm{Mpc})^2}\left(|\tilde H_+(x;q,\hat n)|^2+|\tilde H_\times(x;q,\hat n)|^2\right).
\end{equation}
Next, we define
\begin{equation}
    \mathcal{J}(x;q,\hat n)\equiv\left(\frac{c^2T_\odot^4}{\mathrm{Mpc}^2}\right)\left(|\tilde H_+(x;q,\hat n)|^2+|\tilde H_\times(x;q,\hat n)|^2\right)=\frac{(D/1\,\mathrm{Mpc})^2}{(M/M_\odot)^4}\mathcal{I}(f;\bm\theta),
    \label{eq:scaling}
\end{equation}
where $\mathcal{J}$ only depends on $x,q$ and $\hat n$. To simplify the notation, from now on we measure masses in units of $M_\odot$ and distances in Mpc, i.e., $M$ and $D$ denote the dimensionless numbers $M/M_\odot$ and $D/{\rm Mpc}$.
The physical significance of $\mathcal{J}(x;q,\hat n)$ is then the intensity of a \ac{GW} generated by a binary with total mass $1\,M_\odot$ at distance $D=\,1\mathrm{Mpc}$, with $x=MfT_\odot$.
\end{widetext}

Averaging over the orientation $\hat n$, we define
\begin{equation}
J(x;q)\equiv \langle \mathcal{J}(x;q,\hat n)\rangle_{\hat n}.    
\end{equation}
Depending on the waveform model, this average can be computed exactly, either analytically or by an exact quadrature that requires only a finite number of waveform evaluations; technical details are given in App.~\ref{app:average_over_n}.

We highlight that the orientation-averaged intensity $\langle\mathcal{I}(f;q,\hat n)\rangle_{\hat n}$ of a binary of \emph{any mass} at \emph{any distance} then follows from this \emph{single} two-dimensional table $J(x;q)$ by the rescaling shown in Eq.~\eqref{eq:scaling}. Moreover, because $J$ is independent of $M$ and $D$, the table can be built from waveforms generated at any convenient reference mass $M_{\rm ref}$ and distance $D_{\rm ref}$. In Table~\ref{tab:grids} we list the default values of $M_{\rm ref}$ adopted in \texttt{StochAXtic} for each class of \ac{CBC}, while  $D_{\rm ref}$ is fixed to 1\,Mpc .

The discussion so far does not consider the expansion of the Universe, but assumes a static, flat background spacetime. Nonetheless, for a binary at cosmological redshift $z$, the observed waveform takes exactly the same form once the total mass is replaced by its detector-frame value, $M\to M_{\det}=(1+z)M$, and the distance by the luminosity distance, $D\to d_L(z)$. No new table is therefore needed.

Based on above discussion, we can now rewrite Eq.~\eqref{eq:integral_stage0} as
\begin{widetext}
\begin{equation}
\begin{aligned}
    \OGW(f)&=\frac{4\pi^2f^3}{3H_0^2}\int\D z\frac{R(z)}{1+z}\frac{\D V_c}{\D z}\int \D m_1\D m_2 \,p(m_1,m_2)\left\langle \mathcal{I}\left(f;m_1(1+z),\,m_2(1+z),\,d_L(z)\right)\right\rangle_{\hat n}\\
    &=\frac{4\pi^2f^3}{3H_0^2}\int\D z\frac{R(z)}{1+z}\frac{\D V_c}{\D z}\int \D m_1\D m_2 \,p(m_1,m_2)\frac{M_{\det}^4}{d_L^2} J(x;q),
\end{aligned}
    \label{eq:integral_stage1}
\end{equation}
\end{widetext}
and introduce several numerical tricks we implement in \texttt{StochAXtic} in order to fully carry it out in real time ($\sim\mathcal{O}(\mathrm{ms})$).

\paragraph{Changing variables} Instead of integrating over $m_1,\,m_2,\,z$ directly, we work in log-space: $(\ell,\,\lambda,\,\xi)\equiv(\ln M,\,\ln q,~\ln(1+z))$. 

Changing variables from $(m_1,\,m_2)$ to $(\ell,\,\lambda)$ gives the Jacobian
\begin{equation}
\left|\frac{\p(m_1,\, m_2)}{\p(\ell,\,\lambda)}\right|=\frac{M^2q}{(1+q)^2}.
\end{equation}
Therefore, we can define
\begin{equation}
P(\ell,\lambda)\equiv p(m_1,\,m_2)\frac{M^2q}{(1+q)^2}.
\end{equation}
We have $\D z=(1+z)\D \xi$, so we can rewrite Eq.~\eqref{eq:integral_stage1} as follows:
\begin{widetext}
\begin{equation}
\OGW(f)=\frac{4\pi^2f^3}{3H_0^2}\int\D\xi\int\D\ell\int\D\lambda\,\rho(\xi)P(\ell,\,\lambda)M_{\det}^4J(M_{\det} fT_\odot;\lambda),\quad \rho(\xi)\equiv \frac{R(z)}{d_L^2}\frac{\D V_c}{\D z}\Big|_{\xi=\ln(1+z)}.
\end{equation}
If we further define $\ell_{\det}\equiv\ln M_{\det}=\ln M(1+z)=\ell+\xi$, the above integral reduces to
\begin{equation}
\begin{aligned}
\OGW(f)&=\frac{4\pi^2f^3}{3H_0^2}\int\D\xi\int\D\ell\int\D\lambda\,\rho(\xi)P(\ell,\lambda)e^{4\ell_{\det}}J(e^{\ell_{\det}}fT_\odot;\lambda)\\
&=\frac{4\pi^2f^3}{3H_0^2}\int\D\xi\int\D\ell_{\det}\int\D\lambda\,\rho(\xi)P(\ell_{\det}-\xi,\lambda)e^{4\ell_{\det}}J(e^{\ell_{\det}}fT_\odot;\lambda).
\end{aligned}
\end{equation}
\end{widetext}
From the first line to the second line, we use the fact that the change of variables from $(\ell,\,\xi)$ to $(\ell_{\det},\xi)$ has unit Jacobian.

We change the order of the integral to integrate over $\xi$ first:
    \begin{widetext}
    \begin{equation}
    \begin{aligned}
\OGW(f)&=\frac{4\pi^2f^3}{3H_0^2}\int\D\ell_{\det}\int\D\lambda\, e^{4\ell_{\det}}J(e^{\ell_{\det}}fT_\odot;\lambda)\int\D\xi\,\rho(\xi)P(\ell_{\det}-\xi,\lambda)\\
&\equiv \frac{4\pi^2f^3}{3H_0^2}\int\D\ell_{\det}\int\D\lambda\, e^{4\ell_{\det}}J(e^{\ell_{\det}}fT_\odot;\lambda)K(\ell_{\det},\lambda),
    \end{aligned}
    \label{eq:integral_stage2}
    \end{equation}
    where we have defined
    \begin{equation}
        K(\ell_{\det},\lambda)\equiv \int\D\xi\,\rho(\xi)P(\ell_{\det}-\xi,\lambda).
    \end{equation}
    \end{widetext}
    \begin{center}
\begin{table*}[tp]
\centering
\renewcommand{\arraystretch}{1.3}
\begin{tabular}{ccc}
\hline\hline
Operation & Definition & Fourier domain \\
\hline
Convolution & $(a*b)_k=\sum_d a_d\,b_{k-d}$ & $\mathrm{FFT}(a*b)=\tilde a\,\tilde b$ \\
Cross-correlation & $(a\star b)_k=\sum_d a_d\,b_{k+d}$ & $\mathrm{FFT}(a\star b)=\overline{\tilde a}\,\tilde b$ \\
\hline\hline
\end{tabular}
\caption{Discrete convolution and cross-correlation and their Fourier-domain
forms, with $\tilde a=\mathrm{FFT}(a)$ and $\overline{\tilde a}$ the complex conjugate of
$\tilde a$. Indices are taken modulo the array length, and the
cross-correlation relation assumes a real $a$.}
\label{tab:conv_corr}
\end{table*}
\end{center}

\paragraph{Same spacing of the grid} We now cast the above integral in terms of discrete summations. We use the \emph{same} spacing $\delta\equiv 0.005$ in the log-space for the log-source-frame mass $\ell$, log-redshift $\xi$, log-detector-frame mass $\ell_{\det}$, log-frequency $\ln f$, and $\ln x$. This choice simplifies the calculation significantly, as given a source-frame total mass $\ell_i=\ell_0+i\delta$ and redshift $\xi_j=\xi_0+j\delta$, one can immediately obtain $\ell_{\det, d}=\ell_i+\xi_j=\ell_0+\xi_0+d\delta$, with $d\equiv i+j$. We also have $\ln x=\ln T_\odot+\ell_{\det}+\ln f$, and therefore the index of $\ln x$ is entirely determined by the index of $\ell_{\det}$ and the index of $\ln f$, i.e., $(\ln x)_t=\ln T_\odot+\ell_{\det,d}+(\ln f)_k$. Since $\ln x$,  $\ell_{\det}$, and $\ln f$ all have identical spacing, we then have the rather convenient relation $t = d+k$.

\paragraph{Acceleration via \ac{FFT}} We now consider $K(\ell_{\det,d},\lambda_s)$:
\begin{equation}
    K(\ell_{\det,d},\lambda_s)\equiv K_{d,s}=\delta\sum_j \rho_jP_{d-j,s}\equiv \delta(\rho*P)_{d,s},
        \label{eq:convolution}
\end{equation}
where $\rho*P$ denotes the \emph{convolution} between $\rho$ and $P$. We emphasize again that it is because we choose the same $\delta$-spacing for $\ell, \xi$ and $x$, that we can write the index of $P(\ell_{\det,d}-\xi_{j},\lambda_s)$ as $d-j$.

From the above expression we see that the mass ratio, $\lambda$, is not affected by the redshift, and does not enter the convolution. As a result, a different spacing of $\lambda$ and a different quadrature rule can be adopted to decrease the numerical cost. As we will shortly see that each $\lambda$ node costs one column of the precomputed $J$ table, we keep $n_q$ small and compensate with a higher-order rule: Simpson's rule, whose error is at $\mathcal{O}(\Delta^4)$, while a trapezoid rule would be only $\mathcal{O}(\Delta^2)$, where $\Delta$ is the step size.

In \texttt{StochAXtic}, we adopt the following spacing for $\lambda$:
    $$\Delta_q=\frac{\ln(1/q_{\min})}{n_q-1},\quad\lambda_s=\ln q_{\min}+s\Delta_q,$$
    and we have
    \begin{equation}
        \OGW(f_k)=\frac{4\pi^2f_k^3}{3H_0^2}\delta^2\sum_s \varpi_s \sum_de^{4\ell_{\det,d}}K_{d,s}J_{d+k,s},
        \label{eq:summation_stage1}
    \end{equation}
where $\varpi_s$ are Simpson's weights:
\begin{equation}
\{\varpi_s\}_{s=1}^{n_q}=\frac{\Delta_q}{3}\times\Big\{1,4,2,4,\cdots,2,4,1\Big\}.
\label{eq:varpi}
\end{equation}

For a given $s$, we can now group $e^{4\ell_{\det,d}}$ and $K_{d,s}$ to define \begin{equation}
    A_{d,s}\equiv e^{4\ell_{\det, d}}K_{d,s},
\end{equation}
and hence
\begin{align}
\OGW(f_k)&=\frac{4\pi^2f^3_k}{3H_0^2}\delta^2\sum_{s}\varpi_s\sum_d A_{d,s}J_{d+k,s}\nonumber\\
&\equiv\frac{4\pi^2f_k^3}{3H_0^2}\delta^2\sum_s \varpi_s(A\star J)_{k,s},
\label{eq:cross-correlation}
\end{align}
in which $A\star J$ denotes the \emph{cross-correlation} between $A$ and $J$. 

We note that Eq.~\eqref{eq:convolution} and Eq.~\eqref{eq:cross-correlation} correspond to the convolution between $\rho$ and $p$, and the cross-correlation between $A$ and $J$, and both can be carried out via the \ac{FFT}: see Table~\ref{tab:conv_corr} for details.
\paragraph{Truncated \ac{SVD}}:
Given Eq.~\eqref{eq:cross-correlation}, the calculation of $\OGW(f_k)$ reduces to pre-computing a $J(x;\lambda)$ table beforehand, and calculating two \ac{FFT}'s later. 
One can further accelerate the calculation by decreasing the dimensionality of the $J(x;\lambda)$ table via truncated \ac{SVD}.
Instead of decomposing $J$ directly, we define a matrix $\mathsf{J}_{ts}$ and then perform truncated \ac{SVD} for numerical accuracy (see App.~\ref{app:SVD} for technical details):
\begin{equation}
\mathsf{J}_{ts}\equiv\sqrt{\varpi_s}\frac{J(x_t;\lambda_s)}{\varsigma(x_t)},\quad t=1,...,n_x,\quad s=1,...,n_q,
\end{equation}
where $\varsigma(x)$ is the root-mean-square of $J(x;\lambda)$ over $\lambda$. 

Its \ac{SVD} is then:
\begin{equation}
    \mathsf{J}=\mathsf U\mathsf{\Sigma}\mathsf V^{\top},
    \quad\mathsf J_{ts}=\sum_{r=1}^m \mathsf{U}_{tr}\sigma_r\mathsf{V}_{sr},\quad m=\min(n_x,n_q),
\end{equation}
where $\sigma_r$ is the $r^\mathrm{th}$ singular value, $\mathsf{\Sigma}=\mathrm{diag}(\sigma_1,\cdots,\sigma_m),$ with $\sigma_1\geqslant \sigma_2\geqslant \cdots\geqslant 0$, and
$\mathsf U\in\mathbbm{R}^{n_x\times m}$ and $\mathsf V\in\mathbbm{R}^{n_q\times m}$ have orthonormal columns, i.e., $\sf UU^\top=VV^\top=\mathbbm{1}$. For fixed $r$, $\mathsf{U}_{tr}$ is the $r^{\rm th}$ column of $\sf U$, the $r^{\rm th}$ left singular vector, and $\mathsf V_{sr}$ is the $r^\mathrm{th}$ column of $\mathsf V$, the $r^{\rm th}$ right singular vector. 

Keeping only the first $R$ terms with the largest singular values, we then have
\begin{equation}
    J_{ts}\simeq\sum_{r\leqslant R}C_{rs}B_{tr},
\end{equation}
where $B_{tr}\equiv\varsigma_t\sigma_r\mathsf U_{tr}$, and $C_{rs}\equiv \mathsf V_{sr}/\sqrt{\varpi_s}$. 
The $r^{\rm th}$ column of $B$ is the $r^{\rm th}$ basis mode of the table, a function of $x$ alone, and $C_{rs}$ is its coefficient at the mass ratio node $\lambda_s$, so that each column of the table is a linear combination of the same basis modes.

Here we choose $R$ as the number of singular values with $\sigma_j/\sigma_1>T$, i.e., the largest $r$ satisfying this condition, with $T=10^{-4}$ the default threshold of \texttt{StochAXtic} (in Sec.~\ref{sec:convergence} we compare the resulting $\OGW(f)$ when more modes are retained).

Given $B_{tr}$ and $C_{rs}$, one can rewrite Eq.~\eqref{eq:summation_stage1} as
\begin{equation}
    \OGW(f_k)=\frac{4\pi^2f_k^3}{3H_0^2}\delta^2\sum_r\sum_d e^{4\ell_{\det},d}\hat K_{dr}B_{d+k,r},
    \label{eq:cross-correlation_rewrite}
\end{equation}
where we have defined
\begin{equation}
    \begin{aligned}
        \hat K_{dr}=\delta\sum_j\rho_j\hat P_{d-j,r},\quad \hat P_{d-j,r}=\sum_s \varpi_sP_{d-j,s}C_{rs}.
    \end{aligned}
    \label{eq:hat_K_hat_P}
\end{equation}
For the case considered here, in which only mass and redshift distributions are varied, this reduces the number of \ac{FFT}-based convolutions from $n_q\simeq\mathcal{O}(10^2)$ to $R\simeq \mathcal{O}(10)$.
Since the $J_{ts}$ table does not depend on the population, the truncated \ac{SVD}, and hence  $B_{tr}$ and $C_{rs}$, can be computed once, before a population model is chosen. 
Once an arbitrary population is specified, \texttt{StochAXtic} evaluates $\OGW(f)$ in three steps:
\begin{enumerate}
    \item Evaluate the population density $P_{d-j,s}$ on the grid and combine it with basis coefficient to form $\hat P_{d-j}$ (Eq.~\eqref{eq:hat_K_hat_P});
    \item  Convolve $\hat P_{d-j,r}$ with the redshift weight $\rho_j$ to form $\hat K_{dr}$ (Eq.~\eqref{eq:hat_K_hat_P});
    \item  Cross-correlate $\hat K_{
    dr}$ with the basis modes $B_{d+k,r}$ to get $\OGW(f)$ (Eq.~\eqref{eq:cross-correlation}).
\end{enumerate}
In step (1), one needs to evaluate $P_{d-j,s}$ on the grid, which is the population density of binaries with source-frame total mass $M_{\rm src}$ and mass ratio $q$. Many population models contain hard cutoffs, e.g., at $m_{\min}$ or $m_{\max}$, which cause two problems on a grid of spacing $\delta$. First, whether a cell lies inside or outside the cutoff is an all-or-nothing decision, so the error in $\OGW(f)$ is of first order in $\delta$ and depends on where the edge falls within a cell. Second, $\OGW(f)$ becomes a step function of the edge position, whose derivative vanishes almost everywhere; this spoils gradient-based samplers such as \ac{NUTS}~\cite{Hoffman2014NUTS}
whenever the edge is a free parameter. For such models, we address the first problem by averaging the population density over $n_{\rm sub}=4$ equally spaced points within each $\ln M$ cell, and the second by replacing every hard edge with a logistic step of width $\delta/n_{\rm sub}$ in $\ln M$. Together, they make $\OGW(f)$ a smooth function of the edge positions and reduce the error from the edges to well below that of the other numerical steps (see Sec.~\ref{sec:convergence}). Models whose edges are already smoothly tapered over many cells need neither solution, so we use $n_{\rm sub}=1$.

Steps (2) and (3) are done with \acp{FFT}, so the cost is $\mathcal{O}(N\log N)$ on a grid of $N$ points, and no waveform is evaluated in this stage.
All these aforementioned steps are linear, and therefore can be performed within several milliseconds.

We further note that the basis modes $B_{d+k,r}$ span many decades in $x$ ($B\sim J\sim x^{-7/3}$ in the inspiral), and an \ac{FFT} of such an array loses the small values at large $x$ to round-off, which may corrupt $\OGW(f)$ at high frequencies. We therefore adopt a pre-conditioning prescription by multiplying an $x^A_{d+k}$ term by $B_{d+k,r}$. We then perform \ac{FFT} of $x^A_{d+k}B_{d+k,r}$, doing a cross-correlation, and eventually divide the result by $x^A_{d+k}$. Here we choose $A=7/3$. We refer the reader to App.~\ref{app:preconditioning} for details. 

As a side remark, in \texttt{StochAXtic}, the $J(x;\lambda)$ table is computed on a grid in $\ln x$ with spacing $2\delta$, twice as coarse as required by the FFTs. After the SVD, the $R$ basis vectors $B_{tr}$ are interpolated to the spacing $\delta$ with cubic splines, while the row normalization $\varsigma_t$, which spans many orders of magnitude, is interpolated in $\ln\varsigma_t$. This halves the cost and memory of the table at negligible loss of accuracy (see Sec.~\ref{sec:convergence} for a detailed comparison).
\begin{widetext}
  Eventually, we compute the following summation in \texttt{StochAXtic}:
  \begin{equation}
  \boxed{
      \OGW(f_k)=\frac{4\pi^2f_k^3   }{3H_0^2}(f_kT_\odot)^{-A}\delta^2\sum_{r=1}^R\overbrace{\sum_d\underbrace{e^{(4-A)\ell_{\det,d}}\overbrace{\sum_j\rho_j\hat P_{d-j,r}}^{\text{convolution via FFT}}}_{\substack{\text{mass and redshift parts}\\
      \text{(update for different pops)}}}\underbrace{\left(x^A_{d+k}B_{d+k,r}\right)}_{\substack{\text{waveform part}\\ \text{(only calculated once)}}}}^{\text{cross-correlation via FFT}}}
      \label{eq:summation_mz}
  \end{equation}
  \end{widetext}
\subsection{Case II: Mass, redshift, and spin distributions}\label{sec:m_z_s}
We now include the spin distribution and consider two cases: aligned and misaligned spins, the latter allowing for precession. In both cases, we model dimensionless spins $\bm\chi_1$ and $\bm\chi_2$ using the following parameterization:
\begin{equation}
\begin{aligned}
&\bm \chi_1=(a_1\sin t_1\cos\phi_1,a_1\sin t_1\sin\phi_1,a_1\cos t_1),\\
&\bm\chi_2=(a_2\sin t_2\cos\phi_{2},
a_2\sin t_2\sin\phi_{2},a_2\cos t_2).
\end{aligned}
\end{equation}
We denote the spin magnitudes of $\bm{\chi}_1$ and $\bm\chi_2$ by $a_1$ and $a_2$, 
respectively, and the tilt angles between individual spins and $\hat z$ as $t_1$ and $t_2$. The azimuthal angle between $\bm\chi_1$ and $\bm\chi_2$ is denoted by $\phi_{12}$. See Fig.~\ref{fig:binary_angles} and discussion there for details.
\begin{figure*}[!htbp]
    \centering
\includegraphics[width=0.99\linewidth]{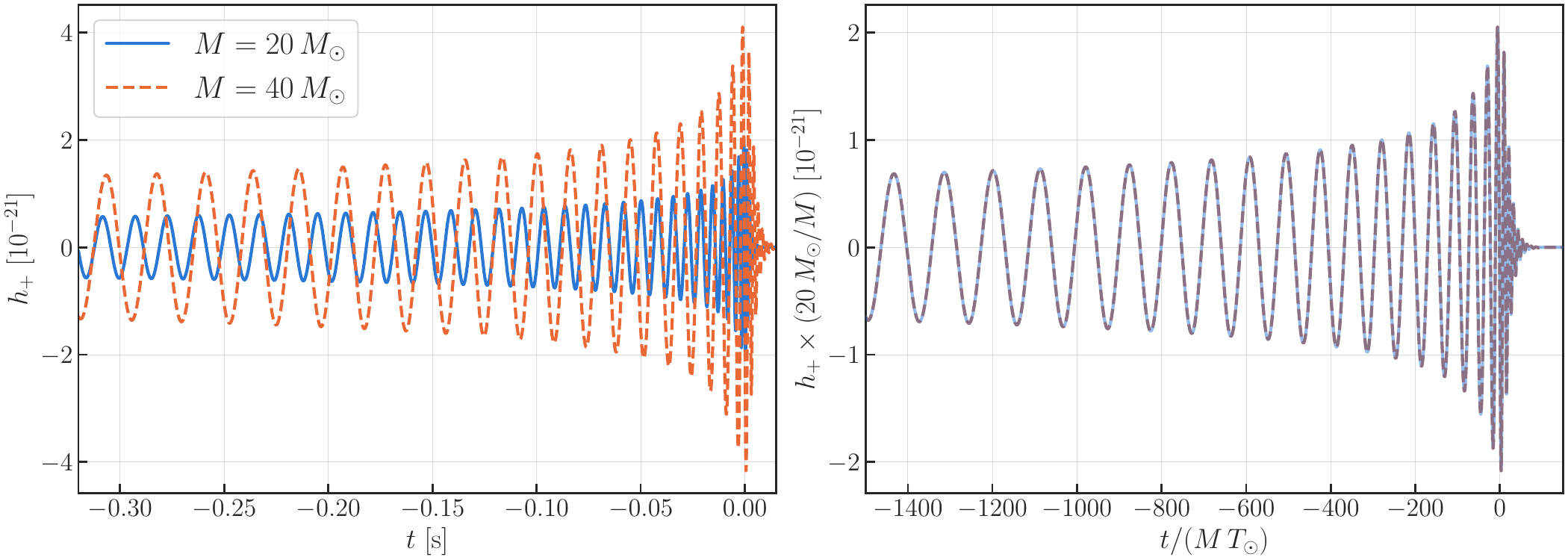}
    \caption{Scale invariance of the waveform. Left: Plus polarization of two binaries that differ only in total mass, $M=20\,M_\odot$ and $40\,M_\odot$, with the same mass ratio $q=0.5$ and aligned spins $\chi_{1z}=\chi_{2z}=0.4$, at the same distance $d_L=1\,\mathrm{Mpc}$ and inclination (IMRPhenomXAS). The heavier binary evolves more slowly and is louder. Right: The same signals with time measured in units of $MT_\odot$ and the strain rescaled by $20\,M_\odot$.}
    \label{fig:scale_td}
\end{figure*}
\subsubsection{Non-precessing case: aligned spins}\label{sec:m_z_s_aligned}
In this sub-case, we assume the two spins are aligned. Consequently, they do not evolve during the inspiral-merger-ringdown process.

We note that the most important pillar of our derivation is the \emph{scale invariance} shown by Eq.~\eqref{eq:pillar}, so whether or not the derivation above holds is determined by whether or not additional physics introduces new physical scales into the problem. Since spins do \emph{not} evolve when they are aligned, it does not matter at which moment they are defined, hence no second physical scale is introduced in the system. Consequently, the total mass $M$ is still the only physical scale, and our derivation in the previous section holds. We show an example in Fig.~\ref{fig:scale_td}, in which we consider two \ac{BBH} systems with the same mass ratio $q=0.5$, same aligned spins $\chi_{1z}=\chi_{2z}=0.4$, but different total masses $20\,M_\odot$ and $40 M_\odot$. 
The left panel shows the time-domain waveforms in SI units, with different frequencies and amplitudes. In the right panel, dividing strain by the total mass $M$ and expressing time in units of $MT_\odot$ makes the two curves coincide, illustrating scale invariance.

Although scale invariance holds when the spins are aligned, in this case, one needs to average $\mathcal{J}(x;\lambda,\hat n,\bm \chi_1,\bm\chi_2)$ over spins $\bm \chi_1$ and $\bm \chi_2$, so the definition of $J(x;\lambda)=\langle \mathcal{J}(x;\lambda,\hat n)\rangle_{\hat n}$ becomes $J(x;\lambda)\equiv \langle \mathcal{J}(x;\lambda,\hat n,\bm \chi_1,\bm \chi_2)\rangle_{\hat n,\bm \chi_1,\bm \chi_2}$. Here, we adopt the reweighting method adopted in \texttt{popstock}~\cite{Renzini:2024pxt}: for each $\lambda$ node we first create a small bank of $J(x;\lambda,\bm \chi_1,\bm \chi_2)$ corresponding to different spins drawn from the proposal distribution $g(\bm\chi_1,\bm\chi_2)$; we then perform importance sampling when a target spin distribution $p(\bm\chi_1,\bm\chi_2)$ is specified later.

The proposal distribution we adopt in \texttt{StochAXtic} is as follows. The tilts and azimuths are drawn uniformly, i.e. $(\cos t_1,\cos t_2)\sim\mathcal{U}[-1,1]^2$, $\phi_1, \phi_{12}\sim\mathcal{U}[0,2\pi)^2$, consistent with the isotropic spin orientation of the population models. For the spin magnitudes, a uniform proposal wastes most samples at large $a$, where the \ac{BBH} population has little support. We therefore use a defensive mixture proposal~\cite{Hesterberg1995} in $r_a=a/a_{\max}$,
\begin{equation}
    g(r_a)=\alpha+(1-\alpha)\mathcal{N}_{[0,1]}(r_a|\mu=0.15,\sigma=0.45),
\end{equation}
where $\alpha=0.3$ and $\mathcal{N}_{[0,1]}(\cdot|\mu=0.15,\sigma=0.45)$ is a truncated normal distribution between $0$ and $1$ with mean $0.15$ and standard deviation $0.45$, which resembles the GWTC-5 spin magnitude distribution~\cite{LIGOScientific:2026ctl}. The uniform component guarantees the importance weights stay bounded for any population, since we have $p(\bm\chi_1,\bm\chi_2)/g(\bm\chi_1,\bm\chi_2)\leqslant p(\bm\chi_1,\bm\chi_2)/\alpha$. The truncated Gaussian component concentrates the samples where \ac{BBH} spins actually lie based on the latest population inference conducted by the \ac{LVK} Collaboration~\cite{LIGOScientific:2026ctl} on GWTC-5.0~\cite{LIGOScientific:2026wfs}. Here we do not use the maximum likelihood values of $\mu\simeq0.06$ and $\sigma\simeq0.36$~\cite{LIGOScientific:2026ctl}, but we use a wider proposal to cover a wider parameter space in $a$.

For \acp{NS}, we restrict the spin magnitudes to $a\leqslant 0.05$, the low-spin prior commonly adopted in \ac{LVK} analyses~\cite{LIGOScientific:2017vwq,LIGOScientific:2018hze}, and we use a uniform proposal.

To sample $g$, we use a low-discrepancy sequence: for each $\lambda$ node we draw $n_\chi$ scrambled Sobol points in the unit hypercube, with an independent scrambling at each node, and map the coordinates of the spin magnitudes through the inverse cumulative distribution of $g$, and those of the tilts and azimuths linearly onto their uniform range. Using such a \ac{QMC} sequence instead of independent random draws speeds up the convergence of the spin average: the error of plain \ac{MC} decreases as $\sim\mathcal{O}(1/\sqrt{N_\chi})$, whereas that of \ac{QMC} can decrease as $\mathcal{O}((\ln n_\chi)^{(d-1)/2}n_\chi^{-3/2})$ for a sufficiently smooth integrand in $d$ dimensions~\cite{Owen1997AoS}. The scrambling keeps the estimator unbiased, and the scatter over independent scrambling provides an estimate of its error (see App.~\ref{app:QMC} for details).

Our goal is to get $\langle J(x;\lambda,\hat n,\bm \chi_1,\bm \chi_2)\rangle_{\hat n,\bm\chi_1,\bm\chi_2}$, where $\hat n$ can be averaged by exactly the same method as we introduced before. The average over spins is given by
\begin{equation}
\begin{aligned}
J(x,\lambda)&\equiv\langle \langle \mathcal{J}(x;\lambda,\hat n,\bm \chi_1,\bm \chi_2)\rangle_{\hat n}\rangle_{\bm s}\\
&=\int\D\bm\chi_1\int\D \bm\chi_2~p(\bm \chi_1,\bm\chi_2)\langle \mathcal{J}(x;\lambda,\hat n,\bm \chi_1,\bm \chi_2)\rangle_{\hat n}.
\end{aligned}
\end{equation}
Leveraging the idea of importance sampling, one can rewrite the above expression as
\begin{equation}
\begin{aligned}
J(x,\lambda)&=\int\D\bm \chi_1\D\bm\chi_2 \frac{p(\bm \chi_1,\bm\chi_2)}{g(\bm \chi_1,\bm\chi_2)}g(\bm \chi_1,\bm\chi_2)\langle \mathcal{J}(x;\lambda,\hat n,\bm \chi_1,\bm\chi_2)\rangle_{\hat n}\\
&\approx\frac{1}{n_\chi}\sum_v^{n_{\chi}}\omega_v\langle \mathcal{J}(x;\lambda,\hat n,\bm \chi_{v,1},\bm\chi_{v,2})\rangle_{\hat n},
\end{aligned}
\end{equation}
where $\omega_v\equiv p(\bm \chi_{v,1},\bm\chi_{v,2})/g(\bm \chi_{v,1},\bm\chi_{v,2})$, and
$p(\bm \chi_{v,1},\bm\chi_{v,2})$ is determined by the specific spin population one considers. In the aligned-spin case, only the components along the orbital angular momentum, $\chi_{1,2;z}=a_{1,2}\cos t_{1,2}$, are retained, while the in-plane components are set to zero; the azimuths $\phi_1$ and $\phi_{12}$ therefore do not enter.

Now, we obtain the expression of $J(x,\lambda)$ again, and we can use the method introduced in the previous case to compute $\OGW(f)$. 

\begin{figure*}[!ht]
    \centering
    \includegraphics[width=0.95\linewidth]{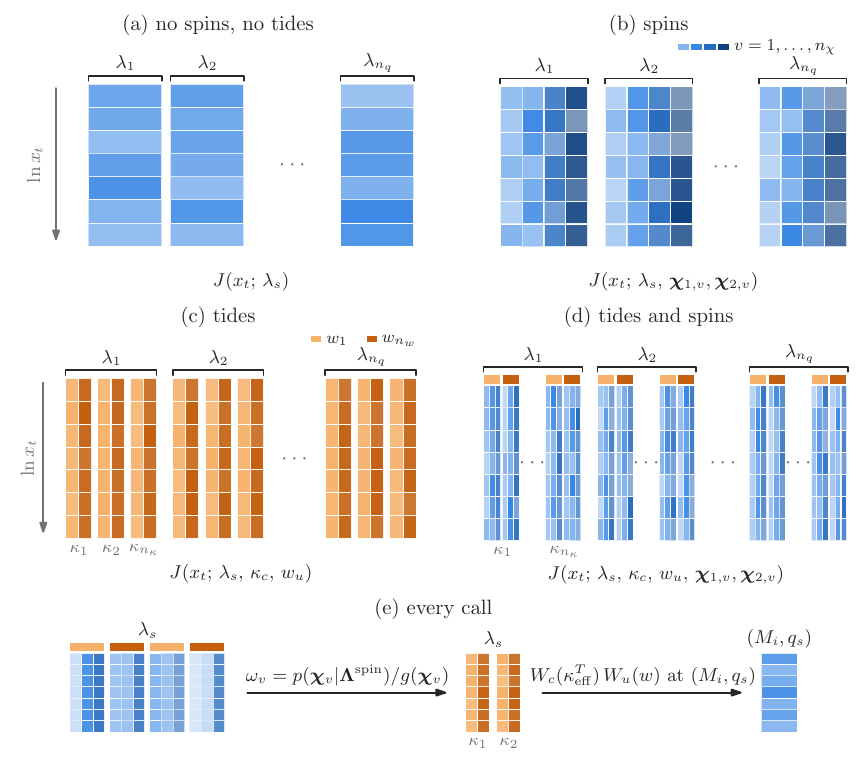}
    \caption{Columns of the $J$ table; the rows are the $\ln x_t$ nodes. (a) Without spins and tides,
one column per $q$ node. (b) With spins, each $q$ node has $n_\chi$ columns, one per spin sample $v$.
(c) With tides, each $q$ node has $n_\kappa\times n_w$ columns, one per node $(\kappa_c, w_u)$.
(d) With both, each tide column splits into the $n_\chi$ spin samples of its $q$ node.
(e) At each evaluation, the spin columns are averaged with the importance weights $\omega_v$,
and the tide columns are interpolated to $\kappa_{\rm eff}$ and $w$ of the binary at $(M, q_k)$,
leaving one column per $(q_k, M)$, as in (a).}
\label{fig:J_table_layer2}
\end{figure*}

In Fig.~\ref{fig:J_table_layer2}, we show a simple sketch of how the $J(x,\lambda)$ table is constructed for various cases. Panel (a) shows the simplest case we discussed in the previous subsection, where only mass and redshift distributions are considered. Panel (b) shows the case where spins are incorporated, where for each $\ln x,\lambda$ node, we have additional $n_\chi$ sub-nodes. When a spin population is specified, one then needs to compute the importance weight $\omega_v$ for each sub-node, and weighted-average them to obtain $J(x,\lambda)$.

In \texttt{StochAXtic}, we perform truncated \ac{SVD} of $\mathsf{J}(x_t;q_s, s_v)\in \mathbbm{R}^{n_x,n_q\times n_\chi}$ to get $B_{tr}$ and $C_{r, (s,v)}$ directly before averaging over spins. 
In this case, the definition of $\mathsf{J}_{t;(s,v)}$ changes to $\sqrt{\varpi_s/n_\chi}J(x_t;\lambda_s,\bm\chi_{v,1}\bm\chi_{v,2})/\varsigma(x_t)$, and we have $C_{r,(s,v)}\equiv\mathsf V_{(s,v),r}\big/\sqrt{\varpi_s/n_\chi}$. Additionally, $\hat P_{d-j,r}$ defined in Eq.~\eqref{eq:hat_K_hat_P} should now be changed to
\begin{equation}
    \hat P_{d-j,r}=\sum_s\varpi_sP_{d-j,s}\bar C_{r,s},\quad \bar C_{r,s}\equiv\frac{\sum_v \omega_{s,v}C_{r,(s,v)}}{\sum_v\omega_{s,v}},
\end{equation}
and Eq.~\eqref{eq:summation_mz} remains valid.

\subsubsection{Precessing case: misaligned spins}\label{sec:m_z_s_precession}
Unlike the aligned-spin case, misaligned spins generally cause the orbital angular momentum $\bm L$ to precess about the total angular momentum $\bm J=\bm L+\bm S_1+\bm S_2$, causing the spins to also evolve. In this case, to guarantee we do not introduce a second physical scale, one has to define spins at a fixed reference $x_\mathrm{ref}=Mf_\mathrm{ref}T_\odot$, rather than a fixed observation frequency, say $20~\mathrm{Hz}$. The default $x_{\rm ref}$ for three types of \acp{CBC} populations are given in Table~\ref{tab:populations2}.

Since the two spins are not aligned, we need to sample $\phi_1$ and $\phi_{12}$ in addition to $\{a_1,a_2,\cos t_1,\cos t_2\}$ when constructing the scrambled Sobol points. Other than that, no calculations in the aligned-spin case need to be modified.

\begin{figure}[!htbp]
    \centering
    \includegraphics[width=0.95\linewidth]{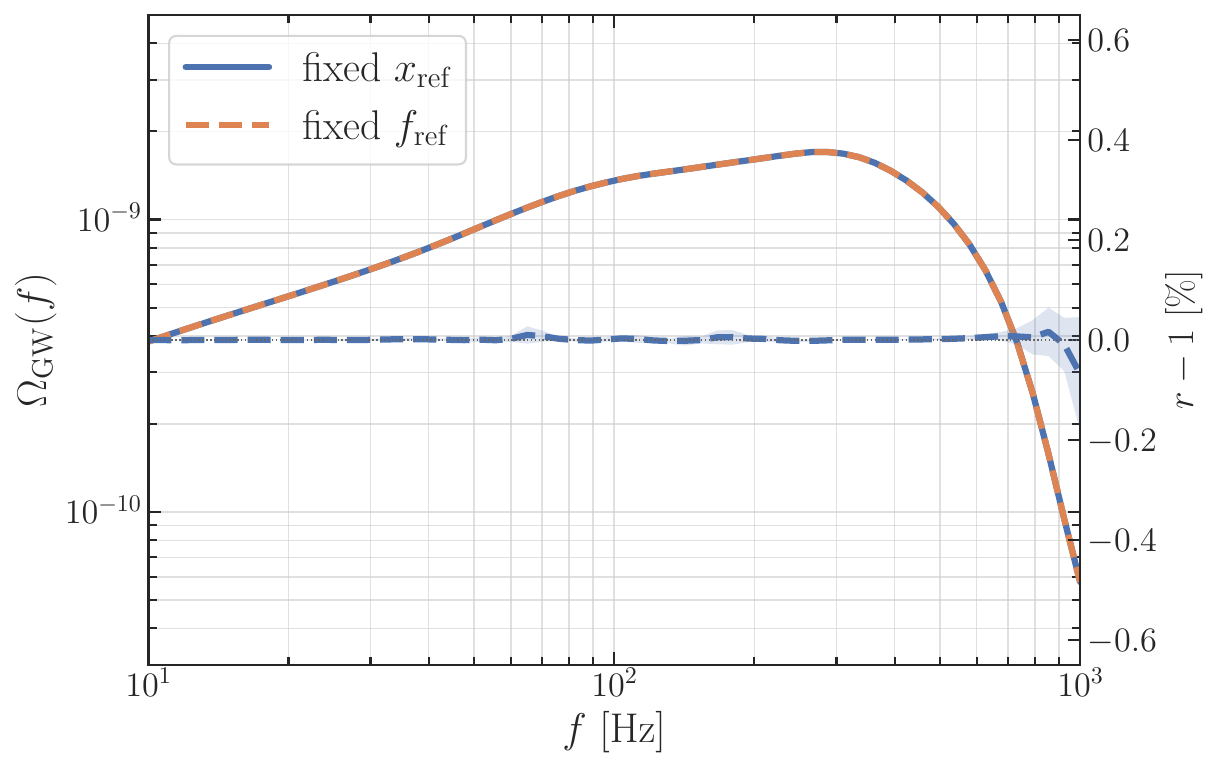}
    \caption{Effect of the spin reference frequency on $\OGW(f)$ for precessing binaries assuming the \texttt{IMRPhenomXPHM} waveform model and the default population models given in Table~\ref{tab:populations} and Table~\ref{tab:populations2}. Left axis: $\OGW(f)$ with the spins defined at a fixed $x_{\rm ref}=0.005$, the convention of \texttt{StochAXtic} (solid blue), and at a fixed detector-frame $f_{\rm ref}=20\,\rm{Hz}$ (dashed orange); the two curves are indistinguishable. Right axis: their relative difference (dashed blue), with a $\pm2\sigma$ band (light blue). Both are computed by direct sampling from the same $2\times10^5$ binaries, so that sampling noise largely cancels in the ratio. The difference is below $10^{-4}$ up to $\sim600$\,Hz and reaches $0.07\%$ at $1$\,kHz, where the merger of the lightest binaries dominates.}
    \label{fig:spin_def_convention}
\end{figure}

We note that the spin population results reported by the \ac{LVK} Collaboration usually define spins at a fixed physical frequency $f_{\rm ref}=20\,\mathrm{Hz}$ instead of a fixed $x_{\rm ref}$. We show in Fig.~\ref{fig:spin_def_convention} that the two definitions only result in minimal differences at $f\gtrsim 10^3\,\mathrm{Hz}$, where not even next-generation detectors are sensitive (cf. Fig.~\ref{fig:convergence}). We stress that both curves in Fig.~\ref{fig:spin_def_convention} are computed from the same $N=2\times 10^5$ binaries drawn from the population assuming that the mass distribution follows a \texttt{Broken Power-Law+Two Peaks} (\texttt{BP2P}) model, the redshift distribution follows a Madau-Dickinson model~\cite{Madau:2014bja}, and the spin distribution follows the default model adopted in the GWTC-5.0 population study~\cite{LIGOScientific:2026ctl}. The choice of population hyperparameters can be found in Table~\ref{tab:populations} and Table~\ref{tab:populations2}; they differ only in the frequency at which the spin vectors are defined. In one case the spins are defined at a fixed detector-frame frequency $f_{\rm ref}=20\,\rm Hz$ for every binary, while in the other they are defined at a fixed $x_{\rm ref}=0.005$, i.e., at $f_{\rm ref}=x_{\rm ref}/(MT_\odot)$, which depends on the mass of the binary. Each binary contributes a term $[\OGW(f)]_i$ to the estimate $\OGW(f)=\sum_i[\OGW(f)]_i/N$, whose standard deviation is $\sigma_\Omega/\sqrt{N}$, with $\sigma_\Omega$ the standard deviation of the $[\OGW(f)]_i$ over the binaries. Since both estimates are computed from the same binaries, their sampling fluctuations are strongly correlated and largely cancel in the ratio $r=\OGW^{f_{\rm ref}}/\OGW^{x_{\rm ref}}$. To first order, the standard deviation of $r$ is that of $(\OGW^{f_\mathrm{ref}})_i-r(\OGW^{x_{\rm{ref}}})$ divided by $\sqrt{N}\OGW^{x_{\rm ref}}$. This is much smaller than the uncertainty of either estimate alone, because the two contributions of each binary are nearly equal. 

One may remove this tiny difference by constructing a transformation to get the spin distribution at a fixed $x_{\rm ref}$ based on \ac{LVK} results, but since this difference is minimal, we leave it for future investigation.
\subsection{Case III: Mass, redshift, spin, and tidal effects}\label{sec:m_z_s_t}
We finally consider the case where NS tidal effects are taken into account. 
\begin{figure*}[!htbp]
    \centering
    \includegraphics[width=0.95\linewidth]{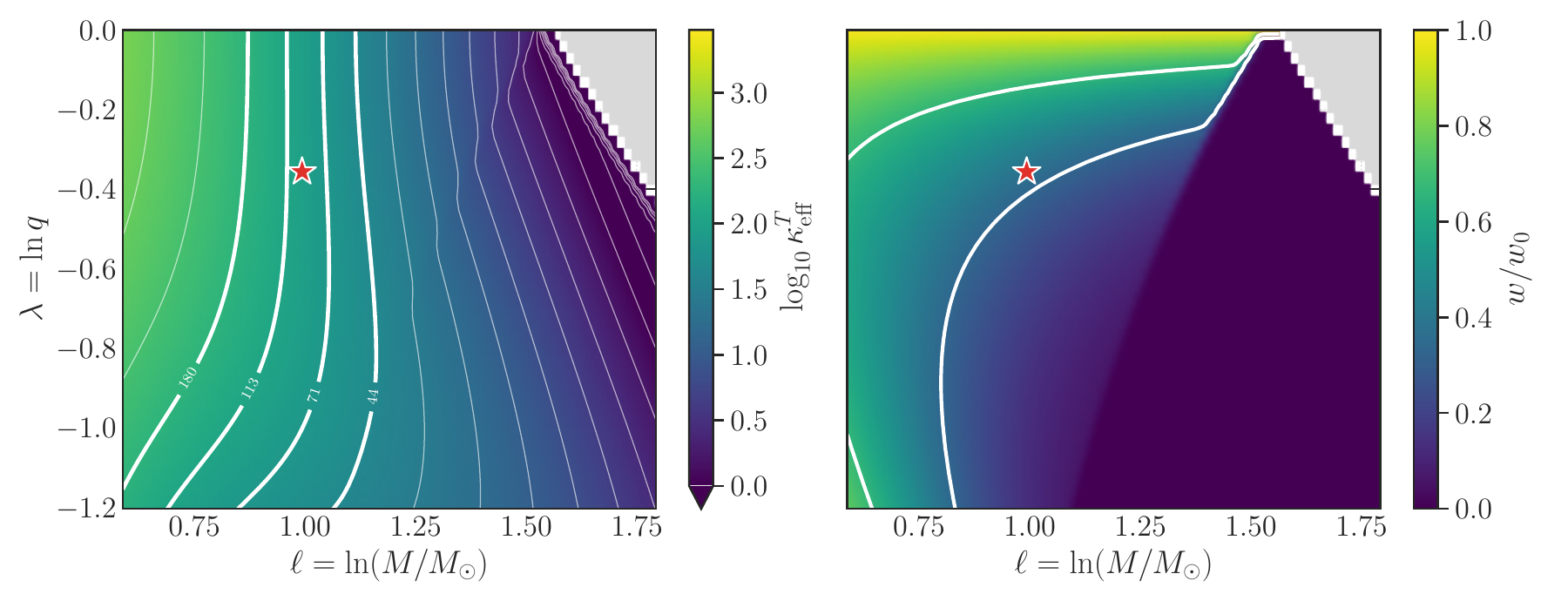}
    \caption{The tidal parameters on the source-frame grid for the default EOS ($\Lambda_{1.4}=400$, $b_1=-6$, $b_2=-3$, $m_{\rm NS,max}=2.3\,M_\odot$). Left: $\kappa_{\rm eff}^T$, with white contours at the nodes $\kappa_c$ of the table; the four nodes used to interpolate at the red star ($M=2.7\,M_\odot$, $q=0.7$, $\kappa_2^T=94.7$) are drawn thick. Right: The share of the primary, $w$, in units of its upper bound $w_0(q)$, with white contours at the interior nodes $w/w_0=1/3$ and $2/3$; the nodes $0$ and $1$ are the ends of the color scale. In the dark region the primary exceeds $m_{\rm NS,max}$, so that $w=0$; in the gray region both stars do, and there are no tides.}
    \label{fig:kappa_demo}
\end{figure*}
Since our target is $\OGW(f)$, which is not affected by the phase of waveforms, we only need to consider how the tidal effects change the waveform amplitude. We denote the tidal deformability of the two objects by $\Lambda_1$ and $\Lambda_2$, respectively, where $\Lambda=0$ for black holes. To the lowest order (5PN)~\cite{Chatziioannou:2020pqz}, $\Lambda_1$ and $\Lambda_2$ only enter the waveform model via their combination, i.e., the dimensionless tidal deformability
\begin{equation}
\tilde\Lambda\equiv\frac{16}{13}\frac{(m_1+12m_2)m_1^4\Lambda_1+(m_2+12m_1)m_2^4\Lambda_2}{M^5}.
\end{equation}
In the \texttt{NRTidal} model~\cite{Dietrich:2017aum} and \texttt{NRTidalv2} model~\cite{Dietrich:2019kaq} one defines the effective tidal coupling constant
\begin{equation}
\begin{aligned}
        \kappa_{\rm eff}^T&=\frac{3}{16}\tilde\Lambda\\
    &\equiv c_1\Lambda_1+c_2\Lambda_2,
\end{aligned}
\end{equation}
which fully captures the amplitude and phase modulations due to tidal effects. In the latest \texttt{NRTidalv3} model~\cite{Abac:2023ujg}, the amplitude modulation is still determined by $\kappa_{\rm eff}^T$, whereas the merger frequency is controlled by another constant, 
\begin{equation}
\begin{aligned}
    \kappa_2^T&=3\frac{m_1m_2}{M^2}\left[\frac{m_1^3}{M^3}\Lambda_1+\frac{m_2^3}{M^3}\Lambda_2\right]\\
    &=a_1\Lambda_1+a_2\Lambda_2.
\end{aligned}
\end{equation}
We now define
\begin{equation}
    \frac{\kappa_2^T}{\kappa_{\rm eff}^T}\equiv\rho_1 w+\rho_2(1-w),
\end{equation}
where $\rho_{A}\equiv a_{A}/c_A,\,A\in\{1,2\}$, $w\equiv c_1\Lambda_1/\kappa_{\rm eff}^T$, and $1-w\equiv c_2\Lambda_2/\kappa_{\rm eff}^T$.
Instead of using $\kappa_{\rm eff}^T$ and $\kappa_2^T$ directly, in \texttt{StochAXtic} we use $\kappa_{\rm eff}^T$ and $w$. 
As a reminder, $w$ here denotes the tidal share, not to be confused with the importance weights $\omega_v$ shown in the previous section. We note that $w$ is bounded between $0$ and $1$. Since the tidal deformability decreases with mass for a given \ac{EOS}, $\Lambda_1\leqslant \Lambda_2$, and hence
\begin{equation}
    0\leqslant w\leqslant w_0(q)=\frac{c_1}{c_1+c_2}.
\end{equation}
The lower bound $w=0$ is reached when the primary has a vanishing tidal deformability, e.g., when $m_1$ exceeds the maximum mass of a \ac{NS}. The upper bound corresponds to $\Lambda_1=\Lambda_2$, which for $q<1$ can only be approached by an \ac{EOS} whose $\Lambda(m)$ is nearly flat between $m_2$ and $m_1$. It depends only on the mass ratio, decreasing from $w_0\to 1$ for $q\to 0$ to $w_0=1/2$ for equal masses. For realistic \acp{EOS}, $\Lambda\propto m^{-n}$ with $n\simeq 4-8$~\cite{Chatziioannou:2020pqz}, so $w\lesssim 1/2$.

In \texttt{StochAXtic}, we adopt the following model:
\begin{equation}
\ln\Lambda(m)=\ln\Lambda_{1.4}+b_1\ln\frac{m}{1.4\,M_\odot}+b_2\ln^2\frac{m}{1.4\,M_\odot},
\label{eq:eos}
\end{equation}
where different \acp{EOS} give different values of $\Lambda_{1.4}$, $b_1$, and $b_2$. This is a logarithmic version of the Taylor expansion of $\Lambda(m)$ around $1.4\,M_\odot$ used in Ref.~\cite{DelPozzo:2013ala,Agathos:2015uaa}. In Tab.~\ref{tab:eos_params}, we fit $\ln\Lambda_{1.4},\,b_1$ and $b_2$ for seven \ac{EOS} models provided in \texttt{LALSimulation}~\cite{lalsuite}.

\begin{table}[t]
\begin{ruledtabular}
\begin{tabular}{lcccccc}
EOS & $M_{\rm max}\,[M_\odot]$ & $\Lambda(1.4\,M_\odot)$ & $\Lambda_{1.4}$ & $b_1$ & $b_2$ & rms [\%] \\
\hline
\texttt{WFF1}      & 2.14 & 153  & 155  & $-6.69$ & $-3.21$ & 3.2 \\
\texttt{APR4\_EPP} & 2.16 & 248  & 252  & $-6.60$ & $-3.06$ & 2.9 \\
\texttt{SLY}       & 2.05 & 313  & 324  & $-7.14$ & $-4.56$ & 5.9 \\
\texttt{ENG}       & 2.25 & 414  & 421  & $-6.32$ & $-2.97$ & 3.0 \\
\texttt{MPA1}      & 2.47 & 513  & 516  & $-5.88$ & $-1.75$ & 1.1 \\
\texttt{H4}        & 2.03 & 897  & 932  & $-6.69$ & $-4.84$ & 7.4 \\
\texttt{MS1B}      & 2.80 & 1409 & 1413 & $-5.54$ & $-1.22$ & 0.5 \\
\end{tabular}
\caption{\label{tab:eos_fit}
Fit of Eq.~(\ref{eq:eos}) to the exact $\Lambda(m)$ of representative EOSs
over $1$--$2\,M_\odot$. 
The EOSs are labeled by their names in
\texttt{LALSimulation}~\cite{lalsuite}, whose tabulated EOSs and Tolman–Oppenheimer–Volkoff (TOV) solver are
used to compute the exact $\Lambda(m)$.
$\Lambda(1.4\,M_\odot)$ is the exact value from the TOV
solution, while $\Lambda_{1.4}$, $b_1$, and $b_2$ are the fitted coefficients.
The last column is the root-mean-square of $\Lambda_{\rm fit}/\Lambda_{\rm exact}-1$ over
the fit range.}
\end{ruledtabular}
\label{tab:eos_params}
\end{table}

Tidal effects enter the waveform through the dimensionless parameters $\{\kappa_{\rm eff}^T,w\}$, so they preserve the scale invariance of Eq.~\eqref{eq:pillar} and can be added to the $J$ table as two additional columns. 

At each mass-ratio ($\lambda$) node we compute the intensity at $n_\kappa$ values of $\kappa_{\rm eff}^T$: $\kappa_{\rm eff}^T=0$, and $n_\kappa-1$ values log-spaced between $0.1$ and $3000$. For \texttt{NRTidalv3}~\cite{Abac:2023ujg}, whose merger frequency depends on $\kappa_{2}^T$, and therefore $w$, each $\kappa_{\rm eff}^T$ node is further split into $n_w=4$ nodes in $w$, equally spaced over the physical range, $w_u=(u/3)w_0(q)$ with $u=0,1,2,3$; for \texttt{NRTidalv2}~\cite{Dietrich:2019kaq}, which depends on $\kappa_{\rm eff}^T$ alone, a single node suffices, so we have $n_w=1$. 
The reason why only a few $w$ nodes is enough stems from the fact that $w$ affects the merger frequency but not the waveform amplitude, so the $J$ table takes exactly identical values before merger $(x<x_{\rm merger})$ for any values of $w$; $w$ only affects $J$ when $x\sim x_{\rm merger}$, and this modification is smooth enough to be captured by only a few $w$ nodes (see App.~\ref{app:n_w} for further discussion).

At each node, the waveform is evaluated with $\Lambda_1=w\kappa_{\rm eff}^T/c_1$ and $\Lambda_2=(1-w)\kappa_{\rm eff}^T/c_2$, so that $\kappa_{\rm eff}^T$ takes exactly its nodal value. The table then becomes $J(x;q,\bm\chi_1,\bm\chi_2,\kappa_{\rm eff}^T, w)\in\mathbbm{R}^{n_x,(n_q\times n_\chi\times n_\kappa\times n_w)}$, and its truncated \ac{SVD} gives coefficients $C_{r,(s,v,c,u)}$, with $c$ and $u$ labeling the $\kappa_2^T$ and $w$ nodes, respectively.

Spins and tides are, however, treated differently once a population is chosen. A spin model is a probability distribution, which we average over with importance weights $\omega_{v}$. The \ac{EOS} model, in contrast, assigns deterministic values of $\kappa_{\rm eff}^T(m_1,m_2)$ and $w$ to each binary, so there is nothing to average over: we must instead evaluate the table at that value.
We therefore first average over spins alone, separately for every $(\kappa_{\rm eff}^T,w)$ node,
\begin{equation}
    \bar C_{r, (s,c,u)}=\frac{\sum_v\omega_{s,v,u}C_{r,(s,v,c,u)}}{\sum_v\omega_{s,v,u}},
\end{equation}
and then interpolate in $\kappa_{\rm eff}^T$ and $w$.
\begin{figure*}[!htbp]
    \centering
\includegraphics[width=0.95\linewidth]{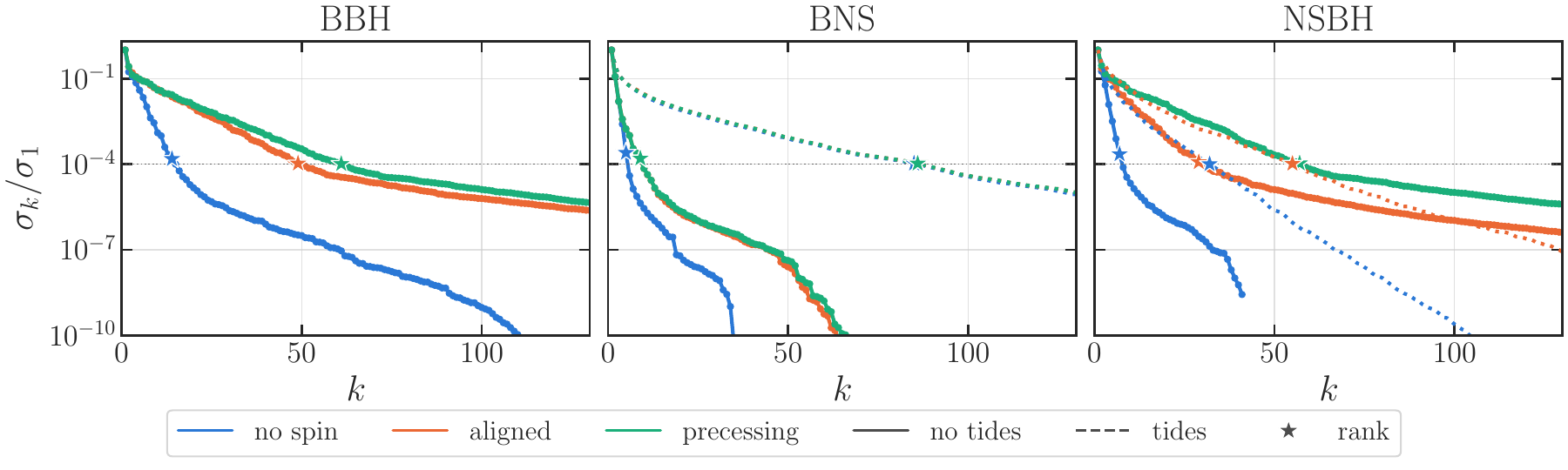}
    \caption{Normalized singular values $\sigma_k/\sigma_1$ of the tables for BBH, BNS and NSBH, without spins, with aligned spins and with precessing spins, without (solid) and with (dashed) tides. Stars mark the rank, the first $k$ with $\sigma_k/\sigma_1<10^{-4}$ (dotted horizontal line).}
    \label{fig:singular_values}
\end{figure*}
The tidal parameters depend on the source-frame masses $m_1$ and $m_2$, and are therefore fixed at every point $(i,s)$ of the source-frame grid (recall that $d\equiv i+j$, so $P_{d-j,s}=P_{i,s}$): the \ac{EOS} gives $\Lambda(m_1)$ and $\Lambda(m_2)$, hence $(\kappa_{\rm eff}^T)_{i,s}$ and $w_{i,s}$. Since both grids are discrete, we interpolate the table between the nodes. 
Because $\kappa_{\rm eff}^T$ spans three decades, we perform cubic Lagrange interpolation in $\ln \kappa_{\rm eff}^T$, using the four nearest log-spaced nodes on the grid. We define the corresponding Lagrange interpolation weights as
\begin{equation}
W^{\kappa}_{i,(s,c)}=\prod_{c’\neq c}\frac{\ln(\kappa_{\rm eff}^T)_{is}-\ln\kappa_{c’}}{\ln\kappa_c-\ln\kappa_{c’}},
\end{equation}
where $c$ and $c'$ run over these four nodes and $W^{\kappa}_{i,(s,c)}=0$ for all other nodes. 
Between the zero-tide node and the first positive log-spaced node, $(\kappa^T_{\rm eff})_1$, we interpolate linearly in $\kappa_{\rm eff}^T$, since its logarithm is undefined at zero.
In $w$, we use cubic Lagrange interpolation on the four nodes $w_u$ of the mass-ratio node $s$, giving us weights:
\begin{equation}
W^{w}_{i,(s,u)}=\prod_{u’\neq u}\frac{w_{is}-w_{u’}}{w_u-w_{u’}}.
\end{equation}
The total weight of the tidal node $(c,u)$ is the product $W_{i,(s,c,u)}=W^{\kappa}_{i,(s,c)}\,W^{w}_{i,(s,u)}$. We note the weights sum to unity in each direction, so the combined weights also sum to unity:
\begin{equation}
\sum_{c,u}W_{i,(s,c,u)}=\Big(\sum_cW^{\kappa}_{i,(s,c)}\Big)\Big(\sum_uW^{w}_{i,(s,u)}\Big)=1 .
\end{equation}
For \texttt{NRTidalv2} there is a single $w$ node and $W^{w}\equiv1$: see panels (c) and (d) of Fig.~\ref{fig:J_table_layer2} for the sketch.

Figure~\ref{fig:kappa_demo} illustrates how the tidal parameters $\kappa_{\rm eff}^T$ and $w$ enter the calculation. For a given \ac{EOS}, every point $(M_{\rm src},q)$ of the source-frame grid has a definite value of $\kappa_{\rm eff}^T$ (left panel) and of the share $w$ of the primary (right panel, in units of its upper bound $w_0(q)$), which are in general not the nodes of the table. Consider the red star $(M=2.7,q=0.7)$, with $\kappa_{\rm eff}^T=94.7$ and $w/w_0=0.38$. For $\kappa_{\rm eff}^T$, it falls between the nodes $\kappa_c=71$ and $113$ (white contours), and its intensity is interpolated from the four nearest nodes, $44,\,71,\,113$ and $180$ (thick contours). For $w$, it falls between the nodes $w/w_0=1/3$ and $2/3$ (white contours), and the four nodes $w/w_0=0,1/3,2/3,1$ are used. Its intensity is thus a weighted sum of $4\times 4$ columns of the table. The two panels also show the physical range of $w$: it reaches its upper bound $w_0$ for equal masses ($q\to 1,$ top), where $\Lambda_1=\Lambda_2$, and its lower bound $w=0$ where the primary exceeds the maximum \ac{NS} mass $m_{\rm NS,\max}$ and has no tidal deformability (dark region). In the gray region, both objects exceed $m_{\rm NS,\max}$ and there are no tides.

Eventually, the definition of $\hat P_{d-j, r}$ should be changed accordingly:
\begin{equation}
    \hat P_{d-j,r}=\sum_s \varpi_sP_{d-j,s}\sum_c\sum_uW_{d-j,(s,c,u)}\bar C_{r,(s,c,u)}.
\end{equation}
Because $\Lambda$ depends on the source-frame mass but not on redshift, $\kappa_{\rm eff}^T$ and $w$ are integrated out on the source-frame grid and before the redshift convolution; all subsequent steps remains unchanged.
\subsection{Singular values}
Before ending this section, we show in Fig.~\ref{fig:singular_values} the singular values of the tables for \ac{BBH}, \ac{BNS} and \ac{NSBH} populations. For \acp{BBH}, a table with only mass and redshift needs $R=14$ modes. Aligned spins raise this to $R=49$, and precession to $R=61$: each spin parameter changes the shape of the spectrum, mainly through the merger and ringdown, and these new shapes need new modes.

For \acp{BNS} without tides, spins raise $R$ only from $5$ to $9$, even with precession, because we assume small spins, $a\leqslant0.05$. Tides raise it to $R=85$–$86$. The \texttt{NRTidal} waveforms are tapered at a merger frequency that depends on $\kappa_{\rm eff}^T$ (Sec.~\ref{sec:m_z_s_t}), so the table contains a sharp cutoff that moves over a wide range of $x$, and many modes are needed to reproduce it.

For \acp{NSBH}, \texttt{IMRPhenomNSBH} supports only aligned spins, and the \texttt{NRTidal} waveforms are restricted to \acp{BNS}, so tides and precession are not combined. Without tides, $R$ grows from $7$ to $29$ with aligned spins and to $57$ with precession; with tides, $R=32$ without spins and $55$ with aligned spins.  As a side note, the solid blue curve terminates at $k=41$, as the $J$ table has only $41$ rows in this case.
\section{Convergence test}\label{sec:convergence}
\begin{figure*}[!htbp]
    \centering
\includegraphics[width=0.95\linewidth]{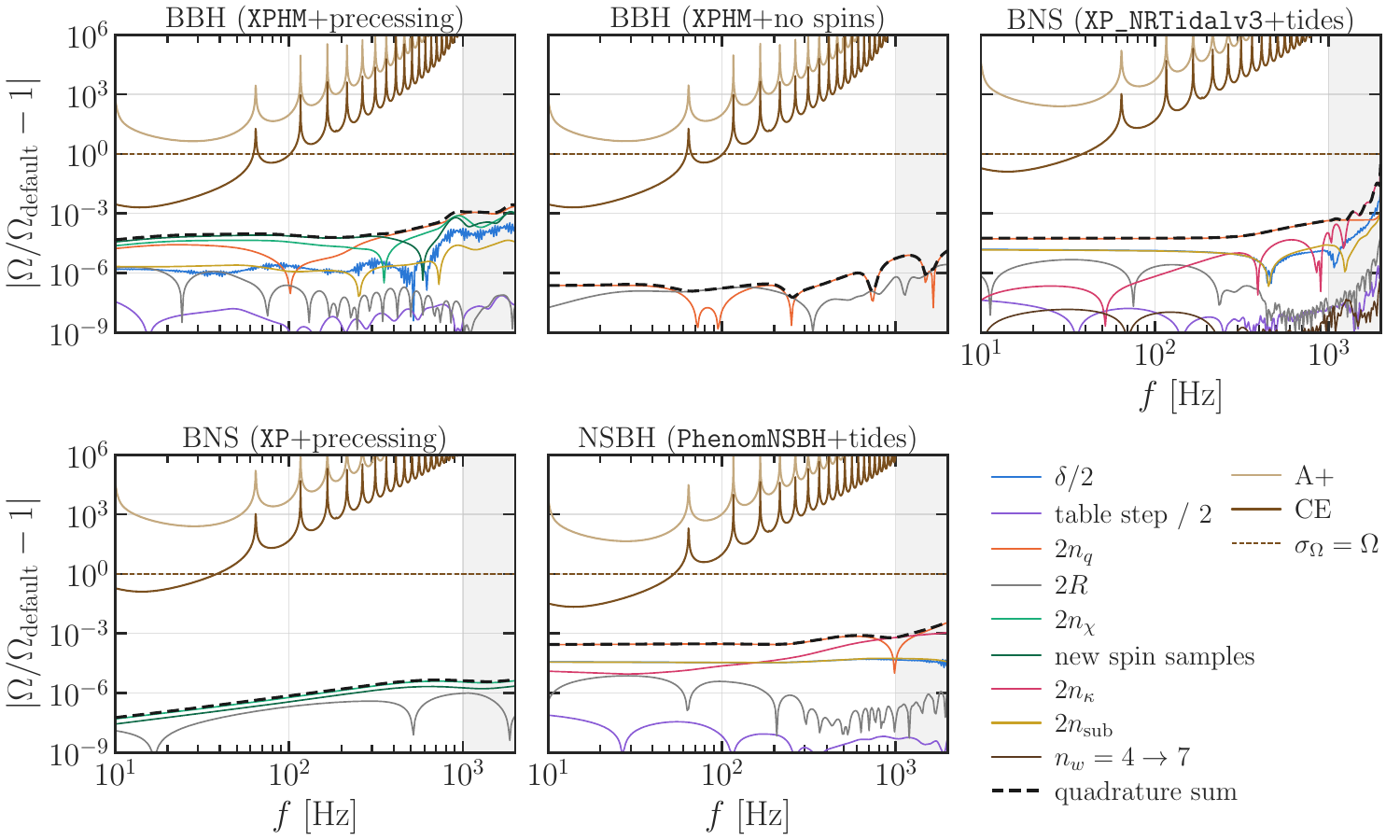}
    \caption{Convergence of \texttt{StochAXtic}. Each curve shows the relative change of $\OGW(f)$ when one resolution parameter is refined from its default value (legend); “new spin samples” redraws the Sobol scrambling. The dashed black curve is their quadrature sum. Brown curves: $\hat\sigma(f)/\OGW(f)$ per $0.25$ Hz bin after one year for the H-L baseline at A+~\cite{O5_PSD_curve} and CE sensitivity~\cite{noise_curves_gwforge}.}
    \label{fig:convergence}
\end{figure*}
In this section we perform a convergence test. In \texttt{StochAXtic}, we have chosen a series of step sizes and a number of nodes. Here, we check how much $\OGW(f)$ changes by doubling the number of nodes or halving the step size. We also compare this difference with the standard deviation of the cross-correlation estimator due to the detector noise. 

We highlight that searches for an \ac{SGWB} currently rely heavily on the cross-correlation method~\cite{Allen:1997ad, Romano:2016dpx,Zhong:2026owg}, where one splits a roughly year-long data stream into shorter segments with duration $\tau$, and defines the cross-correlation estimator $\hat C^{(I)}_f$ for segment $I$:
\begin{equation}
        \hat C_f^{(I)}\equiv \frac{20\pi^2f^3}{3H_0^2\tau}\frac{\real\left[\left(\tilde d_{1;f}^{(I)}\right)^*\tilde d_{2;f}^{(I)}\right]}{\gamma_{12;f}}\;,
\end{equation}
where $\tau$ is the segment length chosen for the analysis, $\gamma_{12;f}$ is the overlap reduction function of the detector baseline~\cite{Allen:1997ad}, and $\tilde d_{a;f},\,a\in\{1,2\}$ denotes the frequency domain strain data for two detectors. We refer readers to Refs.~\cite{Allen:1997ad, Romano:2016dpx} for an overview of how the cross-correlation search works, and to Ref.~\cite{Zhong:2026owg} for a detailed discussion about the statistical properties of this estimator in a practical search. 

One can show that the expected value of this estimator is equal to $\OGW(f)$, and its variance is given by the following expression in the weak-signal limit:
\begin{equation}
  \left(\hat\sigma^{(I)}_f\right)^2 \simeq \left(\frac{20\pi^2f^3}{3H_0^2}\right)^2 \frac{P_{n_{1};f}^{(I)}P_{n_{2};f}^{(I)}} {8\tau\delta\!f\gamma_{12;f}^2}\;,
  \label{eq:var}
\end{equation}
where $\delta\!f$ is the frequency resolution chosen for the Fourier transform, and $P_{n_a;f}^{(I)},\,a\in\{1,2\}$ denotes the detector noise \ac{PSD} in the segment $I$. 

Combining $\hat C_f^{(I)}$ from all available data segments, one gets a final $\hat C_f$ and its corresponding variance $\hat \sigma_f^2$. Ignoring the windowing and overlapping of the data stream and assuming $\hat\sigma^{(I)}_f\simeq\mathrm{const}$, one can get the rough estimate $\hat\sigma^2_f\simeq \left(\hat{\sigma}^{(I)}_f\right)^2/N$. The number of nodes or the step size we choose is sufficient as long as the numerical error in the $\OGW(f)$ computed by \texttt{StochAXtic} is negligible with respect to the detector noise. 

In this work we consider two baselines. As an example, we assume two detectors located at the LIGO Hanford and LIGO Livingston sites, and we assume they have either LIGO A+~\cite{O5_PSD_curve} sensitivity or CE sensitivity~\cite{noise_curves_gwforge} with a total observation time of one year. We also assume $T=4\,\rm{s}$ and $\delta\!f=0.25\,\rm{Hz}$ for the cross-correlation search.

We show the results in Fig.~\ref{fig:convergence}. From the upper left to the lower middle panels, we show five cases: \acp{BBH} using the \texttt{IMRPhenomXPHM} waveform, either including precession (where the spins follow the GWTC-5.0 population model) or without spins; \acp{BNS} using the \texttt{IMRPhenomXP\_NRTidalv3} waveform model including tidal effects; \acp{BNS} using the \texttt{IMRPhenomXP} waveform model including precession (where the spins follow the \texttt{uniform\_isotropic} model, see Table~\ref{tab:populations2}); and \acp{NSBH} using the \texttt{IMRPhenomNSBH} waveform model including tidal effects.

The figure shows that the number of mass ratio nodes, shown in orange, contributes the most to the numerical error, as the total error (dashed black line) falls almost on top of it. Nevertheless, for all five cases the maximum numerical error is at least two orders of magnitude smaller than the detector noise level, and hence negligible.

\section{Simulations}\label{sec:sim}
\subsection{Comparison between \texttt{StochAXtic} and \texttt{popstock}}
Here we compare the $\OGW(f)$ curves computed by \texttt{StochAXtic} with those produced by an \ac{MC} based package, \texttt{popstock}~\cite{Renzini:2024pxt}. Since \texttt{popstock} does not support \ac{NSBH} models, we only compare \ac{BBH} and \ac{BNS} populations. We also ignore spins, as they are not implemented in the current version of \texttt{popstock}.
\begin{figure*}[!htbp]
    \centering
    \includegraphics[width=0.45\linewidth]{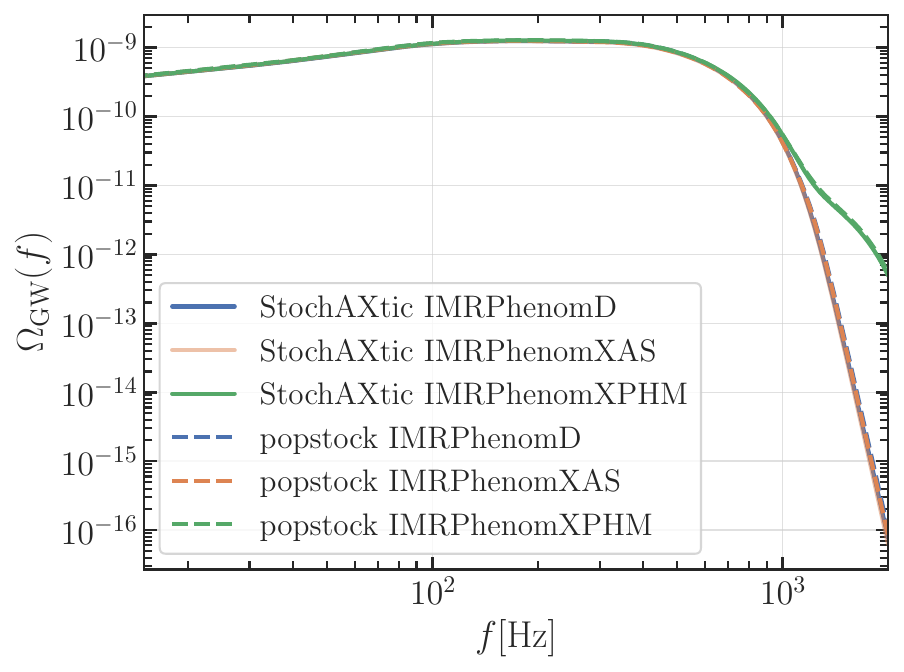}
    \includegraphics[width=0.45\linewidth]{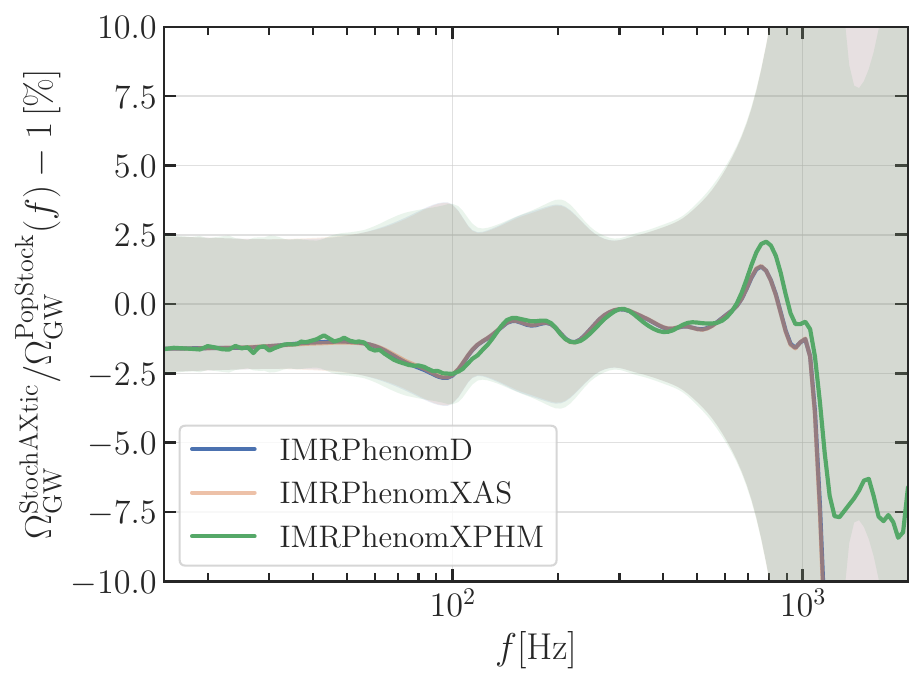}
    \caption{Comparison with \texttt{popstock} for a \ac{BBH} population: Power-Law Plus Peak masses with the default parameters of Table~\ref{tab:populations}, no spins, Madau-Dickinson rate with $R_0=20\,{\rm Gpc^{-3}\,yr^{-1}}$ and $0.2\leqslant z\leqslant10$. Left: $\OGW$ from \texttt{StochAXtic} (solid) and \texttt{popstock} (dashed) for three waveform models. Right: relative difference; the grey band is the $\pm2\sigma$ Monte Carlo uncertainty of \texttt{popstock} with $10^5$ samples. The two codes agree within $0.2$--$0.4\%$ below $100$\,Hz and within the Monte Carlo uncertainty at all frequencies; the band widens at high frequencies, where only the lightest binaries of the sample contribute. Above $\sim1$\,kHz, $\OGW(f)$ from \texttt{IMRPhenomD} and \texttt{IMRPhenomXAS} falls by several decades, while the higher modes of \texttt{IMRPhenomXPHM} keep it larger.}
    \label{fig:BBH_comparison}
\end{figure*}
In Fig.~\ref{fig:BBH_comparison} we show the comparison of \ac{BBH} populations. We assume that the \ac{BBH} mass distribution is described by the \texttt{Power-Law+Peak} (\texttt{PLPP}) model, and that the merger distribution is characterized by the Madau-Dickinson model. In the left panel, we compare three waveform models: \texttt{IMRPhenomD}, \texttt{IMRPhenomXAS}, and \texttt{IMRPhenomXPHM}, with our results shown with solid curves, and those from \texttt{popstock} shown with dashed curves. In the right panel, we show the relative differences between the two packages. We note that \texttt{popstock} is an \ac{MC}-based package, so the result carries \ac{MC} uncertainty. Here, we compute $\OGW^{\rm popstock}(f)$ using $N_{\rm MC}=10^5$ \ac{CBC} events, with each event giving an $\Omega_i(f)$. Suppose that the spread of $\Omega_i(f)$ is $\epsilon(f)$; then the final \ac{MC} uncertainty of $\OGW^{\rm popstock}(f)$ should scale as $\sigma(f)=\epsilon(f)/\sqrt{N_{\rm MC}}$. We show this uncertainty as a shaded band in the right panel. The results show that $\OGW^{\rm popstock}(f)$ and $\OGW^{\rm StochAXtic}$ agree within the statistical uncertainty.
\begin{figure*}[!htbp]
    \centering
    \includegraphics[width=0.45\linewidth]{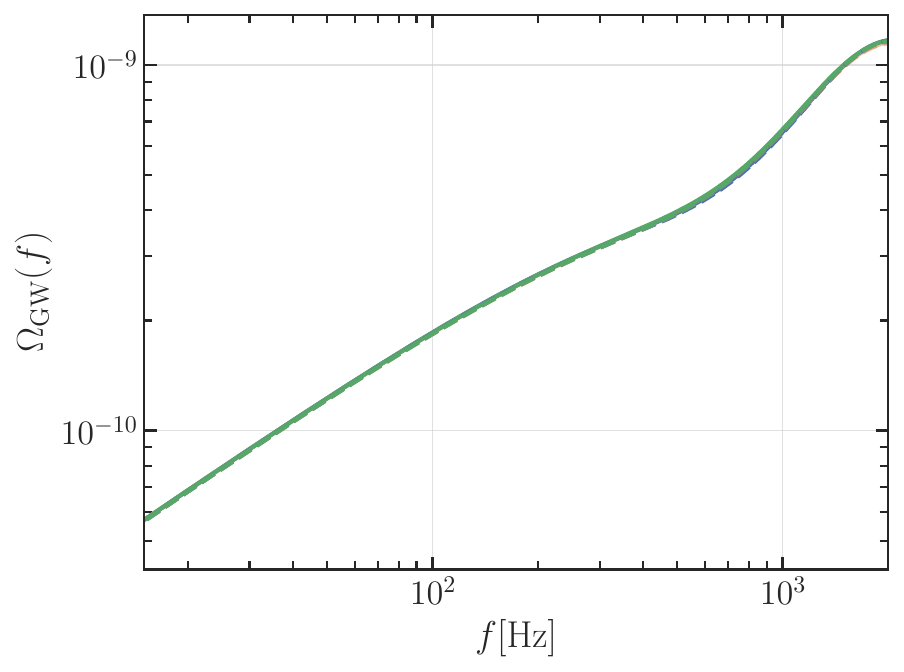}
    \includegraphics[width=0.45\linewidth]{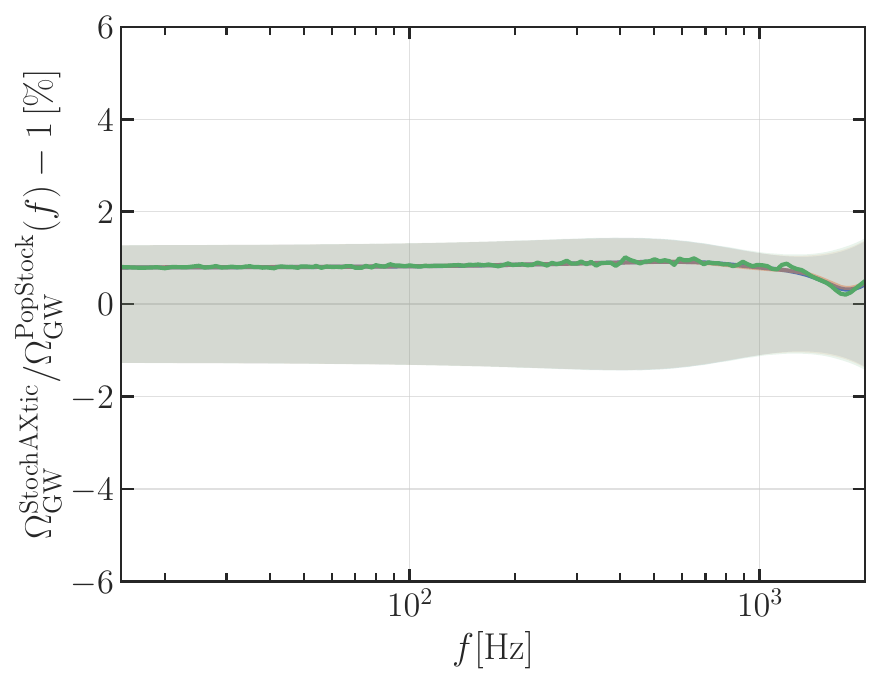}
    \caption{Same as Fig.~\ref{fig:BBH_comparison} for a \ac{BNS} population: component masses uniform in $1\leqslant m_2\leqslant m_1\leqslant2\,M_\odot$, no spins or tides, $R_0=100\,{\rm Gpc^{-3}\,yr^{-1}}$. The relative difference is a constant offset of $1\%$ at all frequencies, within the Monte Carlo uncertainty: all frequencies are computed from the same \texttt{popstock} samples, whose fluctuation shifts the whole curve together. The three waveform models give nearly identical results, because a \ac{BNS} is still in the inspiral phase at $2$\,kHz.}
    \label{fig:BNS_comparison}
\end{figure*}

In Fig.~\ref{fig:BNS_comparison} we show a similar comparison of $\OGW(f)$ for \ac{BNS} populations. We assume a uniform mass distribution model with $m_1\sim\mathcal{U}[1,2]\,M_\odot$, $m_2|m_1\sim\mathcal{U}[1,m_1]\,M_\odot$, and the same Madau-Dickinson merger rate model as the \ac{BBH} population. Fig.~\ref{fig:BNS_comparison} shares the same color codes as Fig.~\ref{fig:BBH_comparison}. We again observe a good consistency of the results given by the two methods.

For both comparisons, we only consider the contribution to $\OGW(f)$ from events with redshift $z\geqslant0.2$, as without this low-redshift cutoff, the results of any \ac{MC}-based methods can be dominated by samples with small $z$ values (see Ref.~\cite{Zhong:2025gxk} for discussion). Here the choice $z_\mathrm{cut}=0.2$ was made merely for comparison, but \texttt{StochAXtic} can support arbitrary $z_{\rm cut}$.
\subsection{A toy model study based on \texttt{StochAXtic} and \texttt{Numpyro}}
In this subsection, we show that since \texttt{StochAXtic} is written in a JAX~\cite{jax2018github}-friendly way, and all the population models are differentiable, one can use it for inferences using \ac{MCMC} methods like \ac{NUTS}~\cite{Hoffman2014NUTS}. 

\begin{figure}[!htbp]
    \centering
    \includegraphics[width=0.99\linewidth]{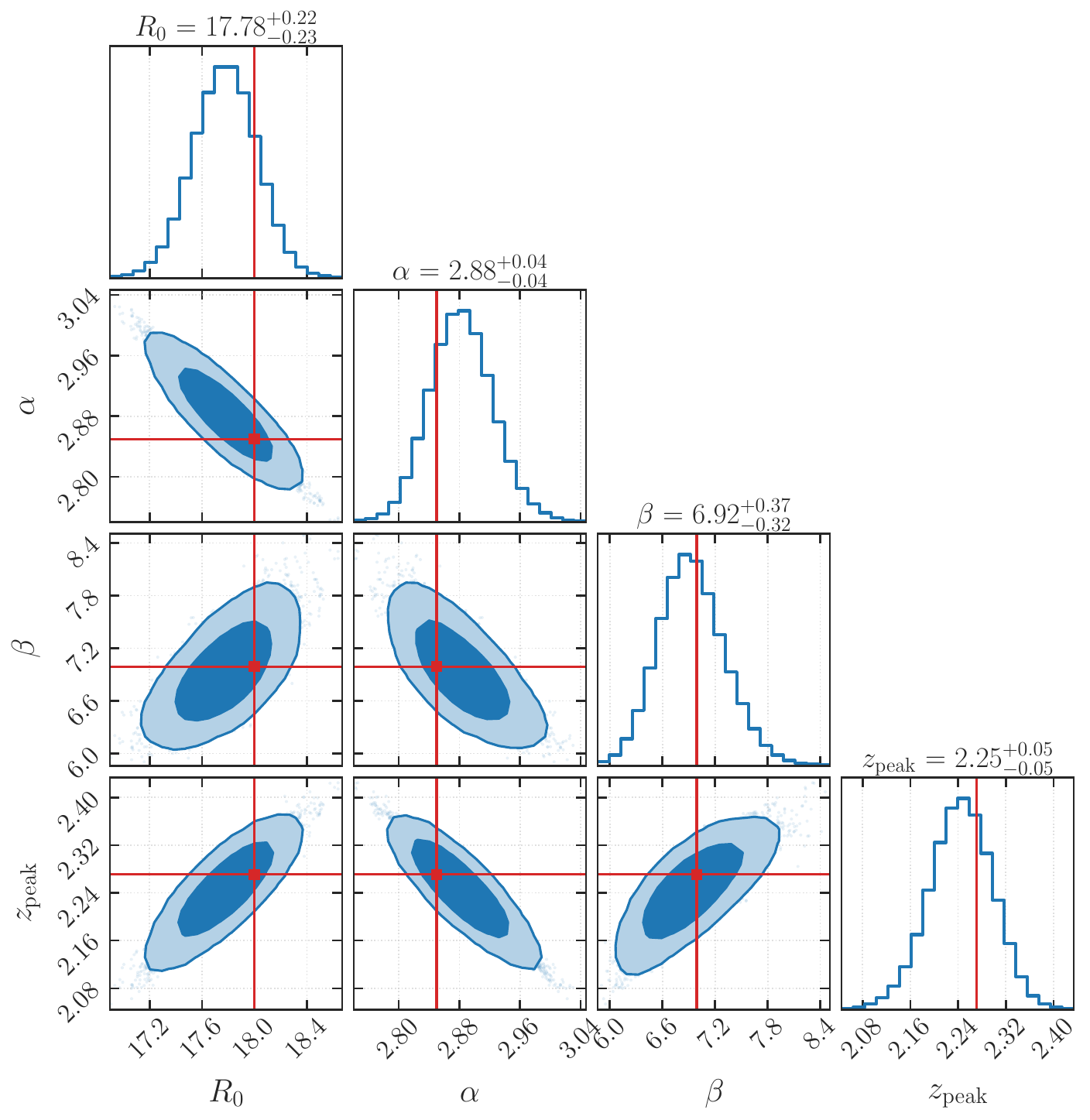}
    \caption{Posterior samples of hyperparameters of the Madau-Dickinson merger rate model. We show 68\% and 95\% credible intervals. The injected true values are shown with red lines.}
    \label{fig:MCMC_demo}
\end{figure}

In this toy example, we consider a \ac{BBH} population without spins, whose mass distribution and redshift distribution are modeled by the \texttt{BP2P} model and Madau-Dickinson model, respectively. We assume we fully understand the mass distribution of \acp{BBH}, and aim to infer the hyperparameters of the redshift distribution, i.e., $R_0,\alpha,\beta$ and $z_{\rm peak}$. We compute $\Omega_\mathrm{BBH}(f)$, and simply assume that the observed spectrum $\Omega_\mathrm{obs}(f)=\Omega_\mathrm{BBH}(f)+\sigma(f)$, where $\sigma(f)$ is chosen to be $\sigma\Omega_{\rm BBH}(f)$, with $\sigma\in\mathcal{N}(0,1)$. We then use the following likelihood to constrain hyperparameters:
\begin{equation}
    -2\ln\mathcal{L}\left(\{\Omega_{\rm obs}\}_{f_0}^{f_{\max}}|\bm\theta_{\rm pop}\right)=\sum_f\frac{(\Omega_{\rm obs}-\Omega_{\rm BBH})^2}{\sigma_f^2}+\rm{const}.
\end{equation}
The priors on the four hyperparameters are $R_0\sim\mathcal{U}[1,50],\,\alpha\sim\mathcal{U}[0,5],\,\beta\sim\mathcal{U}[0,10]$, and $z_{\rm peak}\sim\mathcal{U}[0.5,5
]$.
We use \texttt{Numpyro}~\cite{phan2019composable,bingham2019pyro} to draw posterior samples of the four hyperparameters using the \ac{NUTS} \ac{MCMC} algorithm. The resulting posterior distributions are shown in Fig.~\ref{fig:MCMC_demo}. We successfully recover the four hyperparameters within 68\% credible intervals.

We highlight that although we are not using a realistic $\sigma(f)$, one can easily run the same analysis using $\sigma_f$ values determined by detector noise~\cite{Allen:1997ad, Romano:2016dpx,Zhong:2026owg}. We will use \texttt{StochAXtic} to perform more realistic analyses in the follow-up work~\cite{Zhong:2026}.

\section{Discussion, Outlook, and Conclusions}\label{sec:discussion_conclusion}
In this paper we present \texttt{StochAXtic}, an end-to-end  JAX-accelerated package to compute $\OGW(f)$ for \ac{CBC} populations. Unlike brute-force \ac{MC}-based method, the key pillar of our implementation is the scale invariance of the waveforms: at fixed mass ratio, spins, and tidal deformabilities, the waveform depends on the total mass only through $x=MfT_\odot$. This allows us to move all the expensive waveform evaluations into a preprocessing stage, which computes and caches the table $J(x;\lambda)$ once for all populations. 
A truncated \ac{SVD} further compresses this table into a few basis modes, and only these modes and their coefficients enter the subsequent calculations of $\OGW(f)$. 
Besides the mass and redshift distributions, \texttt{StochAXtic} includes spins and tidal effects, since neither breaks the scale invariance. 
Our convergence test shows that the numerical error of the computed $\OGW(f)$ is below $10^{-3}$ up to $1$\,kHz. This is more than two orders of magnitude below the statistical uncertainty per frequency bin of the cross-correlation estimator assuming one year of observing time for a detector network including two \ac{CE} detectors, and more than four orders of magnitude below that of LIGO detectors assuming $A+$ sensitivity. The results agree with \texttt{popstock} within its statistical uncertainty. Finally, we showed with a toy example how \texttt{StochAXtic} can be used in downstream Bayesian inference.

We close with the limitations of \texttt{StochAXtic} and plans for future work. First, the current version assumes that the mass, redshift, spin, and tidal distributions are independent. Population models with correlations, for example a mass distribution that evolves with redshift or spins that depend on mass, are left for future development. 
Second, in precessing case we define the spins at a fixed $x_{\rm ref}$, whereas the population analyses of the \ac{LVK} Collaboration define them at a fixed detector-frame frequency, usually $f_{\rm ref}=20\,\rm{Hz}$. We showed that this difference has a negligible effect on $\OGW(f)$; a map between the two conventions would remove it altogether.
Third, in Ref.~\cite{Zhong:2026jzl} we showed that a realistic search for the \ac{CBC} background must use the full probability distribution of the cross-correlation estimator, not only its mean $\OGW(f)$, and that this distribution depends on the \ac{CBC} population. \texttt{StochAXtic} computes only the mean, but the numerical techniques developed here (the scale invariance, the precomputed table and its \ac{SVD}) should carry over to that problem.

\section*{Data availability}
\texttt{StochAXtic} and the plotting script used in this work are publicly available at \faGithub~\url{https://github.com/Haowen-Zhong/StochAXtic}.
\begin{acknowledgments}
We thank Francesco Iacovelli for helpful discussions. H.Z., E.V. and E.B. are supported by NSF Grants No.~AST-2606672, No.~PHY-2513337, No.~PHY-090003, and No.~PHY-20043, by the Simons Foundation [MPS-SIP-00001698, E.B.], and by the Simons Foundation International [SFI-MPS-BH-00012593-02]. This project has received funding from the European Union’s Horizon Europe research and innovation programme under the Marie Skłodowska-Curie Staﬀ Exchanges grant agreement No.~101299389 (STRONG). H.Z. is supported by the Postdoctoral Fellowship at Johns Hopkins University. E.V. is supported by the NSF Graduate Research Fellowship Program under Grant No. DGE2631994. Any opinions, findings, and conclusions or recommendations expressed in this material are those of the authors and do not necessarily reflect the views of the NSF.
This work used computational resources provided by the LIGO Laboratory and supported by NSF Grants PHY-0757058 and PHY-0823459. 
\end{acknowledgments}

\appendix
\section{Averages over orientation $\hat n$}\label{app:average_over_n}
\subsection{Nonprecessing waveform models restricted to $(2,\,\pm2)$-modes}\label{app:faceon}
For nonprecessing waveform models containing only the $(2,\,\pm2)$-modes, we have:
\begin{equation}
    \tilde{h}_+(f)=\left(\frac{1+u^2}{2}\right)\tilde {\mathfrak{h}}(f),\quad \tilde h_\times(f)=-iu\tilde {\mathfrak{h}}(f),
\end{equation}
where $u\equiv \cos\iota$, and $\tilde{\mathfrak{h}}(f)$ is the complex frequency-domain waveform shared by both polarizations, with the inclination dependence factored out.
We then have
\begin{equation}
    \mathcal{I}(f,u)=A(u)|\tilde {\mathfrak{h}}(f)|^2,\quad A(u)=\frac{1+6u^2+u^4}{4},
\end{equation}
and $\mathcal{I}(f,u)$ does not depend on $\phi$, so
\begin{equation}
    \langle \mathcal{I}(f,u)\rangle_{\hat n}=\frac{2}{5}\mathcal{I}(f,\iota=0).
\end{equation}
Consequently, for this case, we only need to evaluate $\mathcal{I}(f)$ for the face-on case, i.e., $\iota=0$.
\subsection{Summation of modes}\label{app:modes}
For some waveform models, such as \texttt{IMRPhenomXPHM}~\cite{Pratten:2020ceb}, \texttt{LALSimulation}~\cite{lalsuite} provides the individual modes $\tilde h_{\ell m}$ directly. 
In this case, we can define the complex strain
\begin{equation}
    h(t;\hat n)\equiv h_+(t;\hat n)-ih_\times(t;\hat n)=\sum_{\ell m}h_{\ell m}(t) {}_{-2}Y_{\ell m}(\hat n),
\end{equation}
where ${}_{-2}Y_{\ell m}(\hat n)$ denotes the spin-weighted spherical harmonics. One can show that
\begin{equation}
    \mathcal{I}(f)=|\tilde h_+(f)|^2+|\tilde h_\times(f)|^2=\frac{1}{2}\left(|\tilde h(f)|^2+|\tilde h(-f)|^2\right).
\end{equation}
We then have
\begin{widetext}
\begin{equation}
\begin{aligned}
    \langle \mathcal{I}(f;\hat n)
    \rangle_{\hat n}&=\frac{1}{2}\times\frac{1}{4\pi}\int\D\hat n\left(|\tilde h(f)|^2+|\tilde h(-f)|^2\right)\\
    &=\sum_{\ell m}\sum_{\ell'm'}\left[\frac{\left(\tilde h_{\ell m}(f)\tilde h_{\ell'm'}^*(f)+\tilde h_{\ell m}(-f)\tilde h_{\ell'm'}^*(-f)\right)}{8\pi}\left(\int\D\hat n
    {}_{-2}Y_{\ell m}(\hat n){}_{-2}Y_{\ell' m'}^*(\hat n) \right)\right]\\
    &=\frac{1}{8\pi}\sum_{\ell m}\left(|\tilde h_{\ell m}(f)|^2+|\tilde h_{\ell m}(-f)|^2\right).
\end{aligned}
\label{eq:sum_over_modes}
\end{equation}
\end{widetext}
It is worth noting that the above derivation does \emph{not} assume aligned spins, so one can use this method for waveform models including precession effects.

Before concluding this case, we clarify why the sum in Eq.~\eqref{eq:sum_over_modes} can be evaluated using either inertial frame or co-precessing frame waveform modes when computing $\OGW(f)$.
The co-precessing frame is defined such that its $z$-axis follows the orbital angular momentum $\bm L$ as it precesses about the total angular momentum $\bm J$. 
Precessing waveform models construct the inertial frame modes $\tilde h_{\ell m}$ by rotating the co-precessing modes $\tilde h'_{\ell m}$. If the same rotation is applied to all modes with a given $\ell$, the unitarity of the rotation guarantees that
\begin{equation}
    \sum_m|\tilde h_{\ell m}(f)|^2=\sum_m|\tilde h'_{\ell m}(f)|^2.
    \label{eq:coprec_sum}
\end{equation}
However, as we shall explain below, this equality does not generally hold for individual sources in the frequency domain. We show that it nevertheless holds after averaging over the source population, which is sufficient for computing $\OGW(f)$.

The complication arises because different co-precessing modes are rotated using precession angles evaluated at different orbital frequencies. Specifically, the mode $(\ell,\,m')$ at frequency $f$ is associated with an orbital frequency $f/|m'|$, and the waveform model evaluates the precession angles at the corresponding orbital frequency. Denoting the resulting rotation by $R_{m'}(f)$, the inertial frame modes are given by
\begin{equation}
    \tilde h_{\ell m}(f)=\sum_{m'}D^\ell_{mm'}[R_{m'}(f)]\tilde h_{\ell m'}'(f),
\end{equation}
where $D_{mm'}^{\ell}$ is the Wigner $D$-matrix. Summing the squared amplitudes over $m$ gives
\begin{equation}
\begin{aligned}
 \sum_m |\tilde h_{\ell m}|^2&=\sum_{m'}|\tilde h_{\ell m'}'|^2
    \\
    &+\sum_{m'\neq m''}\left(D^{\ell}[R_{m''}]^\dagger D^\ell[R_{m'}]\right)_{m''m'}\tilde h'_{\ell m'}\tilde h^{\prime *}_{\ell m''}.
\end{aligned}
\label{eq:crossterms}
\end{equation}
The first term is exactly the right-hand side of Eq.~\eqref{eq:coprec_sum}, whereas the second term contains interference between different co-precessing modes with the same $\ell$. If all modes share the same rotation, unitarity makes this term vanish, recovering Eq.~\eqref{eq:coprec_sum}. When the rotations differ, however, the cross terms may not vanish for an individual source.

Crucially, these cross terms vanish upon averaging over the distribution of $\phi_1$. To see this, recall that the spin azimuths $\phi_1$ and $\phi_2$ are defined relative to the line joining the two bodies (see Fig.~\ref{fig:binary_angles}). Consider two binary configurations that only differ by the values of $\phi_1$, with one chosen as the reference configuration at $\phi_1=0$.
By rigidly rotating the configuration with nonzero $\phi_1$ about $\bm L$ by an angle $-\phi_1$, we can align both spin vectors with their counterparts in the reference configuration. 
The two configurations then differ only in the orbital phase, which is shifted by $-\phi_1$. 
Since the co-precessing modes depend on the orbital phase $\Phi$ through $e^{-im'\Phi}$, we obtain
\begin{equation}
    \tilde h'_{\ell m'}(f;\phi_1,\phi_{12})
    = e^{i m'\phi_1}\,\tilde h'_{\ell m'}(f;0,\phi_{12}).
    \label{eq:phi1_phase}
\end{equation}
We note that the matrix $D^\ell[R_{m''}]^\dagger D^\ell[R_{m'}]$ in Eq.~\eqref{eq:crossterms} relates the co-precessing frame at the time when the orbital frequency is $f/|m'|$ to the co-precessing frame at the time when it is $f/|m''|$. It is \emph{not} affected by the rigid rotation described above, and therefore does not depend on $\phi_1$.
Consequently, $\phi_1$ enters Eq.~\eqref{eq:crossterms} only through the co-precessing modes. Using Eq.~\eqref{eq:phi1_phase}, each cross term is the corresponding term at $\phi_1=0$ multiplied by the phase factor $e^{i(m-m'')\phi_1}$, with $m'\neq m''$.
Because $\phi_1$ is assumed to be uniformly distributed, averaging over $\phi_1$ at fixed $\phi_{12}$ eliminates all terms with $m'\neq m''$. Therefore
\begin{equation}
\left\langle\sum_m|\tilde h_{\ell m}(f)|^2\right\rangle_{\phi_1}=
\sum_m|\tilde h'_{\ell m}(f)|^2.
\end{equation}
Thus, the co-precessing modes can be used to evaluate Eq.~\eqref{eq:sum_over_modes} when computing $\OGW(f)$.

This distinction is relevant to our implementation of the waveform models. For \texttt{IMRPhenomXPHM}~\cite{Pratten:2020ceb}, \texttt{LALSimulation}~\cite{lalsuite} provides the inertial frame modes, which we directly use in Eq.~\eqref{eq:sum_over_modes}.
For \texttt{IMRPhenomPv3}~\cite{Khan:2018fmp} and \texttt{IMRPhenomPv3HM}~\cite{Khan:2019kot}, we instead use the co-precessing modes. Constructing the inertial frame modes for these models requires  precession angles that are undefined for a small fraction of spin configurations, causing \texttt{LALSimulation} to return an error. Using the co-precessing modes avoids this numerical issue.

For \texttt{IMRPhenomPv3}, only the $(2,\,\pm2)$ co-precessing modes are present, and only one contributes at positive $f$. There are therefore no cross terms, and Eq.~\eqref{eq:coprec_sum} holds for each individual source. In contrast, \texttt{IMRPhenomPv3HM} additionally includes the $(2,\,\pm1),(3,\,\pm3),(3,\,\pm2),(4,\,\pm4)$ and $(4,\,\pm3)$ modes. Cross terms $(2,\,2)\times (2,\,1)$, $(3,\,3)\times (3,\,2)$, and $(4,\,4)\times (4,\,3)$ can therefore survive for individual sources. Nevertheless, as demonstrated above, these terms vanish after population averaging, so the use of co-precessing modes remains valid for computing $\OGW(f)$. 
\subsection{Most general case}\label{app:general}
For arbitrary waveforms one can numerically integrate $\mathcal{I}(f,\hat n)$ over a grid of $\hat n=(\cos\iota,\phi)$. 
We note that~\cite{Brown:2007jx}
\begin{equation}
    {}_{-2}Y_{\ell m}(\iota, \phi)=y_{\ell m}(\iota)e^{im\phi},\quad y_{\ell m}(\iota)=\sqrt{\frac{2\ell+1}{4\pi}}d_{m,2}^\ell(\iota),
\end{equation}
\begin{widetext}
where $d_{m,2}^\ell$ is the Wigner $d$-function:
\begin{equation}
    d_{m,2}^\ell(\iota)=\sum_k(-1)^{k+m-2}\frac{\sqrt{(\ell+m)!(\ell-m)!(\ell+2)!(\ell-2)!}}{(\ell+2-k)!k!(\ell-k-m)!(k-2+m)!}\cos^{2\ell-2k+2-m}\left(\frac{\iota}{2}\right)\sin^{2k-2+m}\left(\frac{\iota}{2}\right),
\end{equation}
and $k\in\mathbb{Z}$ such that all factorial arguments are non-negative. For example, $d_{2,2}^2=\cos^4\left(\frac{\iota}{2}\right)$, $d^{2}_{-2,2}=\sin^4\left(\frac{\iota}{2}\right),$ and hence ${}_{-2}Y_{2,\pm 2}\propto (1\pm \cos\iota)^2e^{\pm 2i\phi}$.
\end{widetext}
Given this, each term of $\mathcal{I}$ takes the form:
\begin{equation}
    \tilde h_{\ell m}\tilde h_{\ell'm'}^*\underbrace{\left[d_{m,2}^\ell(\iota)d^{\ell'}_{m',2}(\iota)\right]}_{\text{depends on } \iota}\underbrace{\left[e^{i(m-m')\phi}\right]}_{\text{depends on } \phi},
\end{equation}
where $\iota$ and $\phi$ are decoupled, and therefore one can choose different grids for $\iota$ and $\phi$ for the summation.

We first average over $\phi$. Each term of $\mathcal{I}(\phi)$ depends on $\phi$ only through $e^{i(m-m')\phi}$, and since $|m-m'|\leqslant 2L$, where $\mathcal{I}(\phi)$ is a trigonometric polynomial of degree at most $2L$,
\begin{equation}
    \mathcal{I}(\phi)=\sum_{n=-2L}^{2L}c_ne^{in\phi},
\end{equation}
whose average over $\phi$ is then $c_0$. For $K$ equispaced points $\phi_j=2\pi j/K$, we have
\begin{equation}
    \frac{1}{K}\sum_{j=0}^{K-1}e^{in\phi_j}=\begin{cases}
        1,&n\equiv 0\,(\mathrm{mod\,} K),\\
        0,&\mathrm{otherwise}.
    \end{cases}
\end{equation}
Choosing $K=n_\phi=2L+1$ then removes every term with $1\leqslant |n|\leqslant 2L$, and yields the $\phi$ average exactly.

After averaging over $\phi$, we have only $m=m'$ terms, the term depending on $\iota$ become $d_{m,2}^\ell(\iota)d_{m',2}^{\ell'}(\iota)$, and the cosine and sine terms in this product yield
\begin{equation}
    \cos^{2[(\ell+\ell')-(k+k')+2-m]}\left(\frac{\iota}{2}\right)\sin^{2[(k+k')-2+m]}\left(\frac{\iota}{2}\right).
\end{equation}
Changing variables from $\iota$ to $u$, and substituting $\cos^2(\iota/2)$ with $(1+u)/2$, and $\sin^2(\iota/2)$ with $(1-u)/2$, each product becomes a polynomial in $u$ of degree $\ell+\ell'\leqslant 2L$. We observe that the measure $\frac{\D\hat n}{4\pi}=\frac{\D u}{2}\frac{\D \phi}{2\pi}$ is flat with $u\in[-1,1]$, which is exactly the setting of Gauss-Legendre quadrature: an $n_u$-point rule integrates polynomials of degree up to $2n_u-1$ exactly on $[-1,1]$. 
Requiring $2n_u-1\geqslant 2L$ then naturally gives rise to the choice of $n_u=L+1$.

Finally, we have
\begin{equation}
    \langle \mathcal{I}(f,\hat n)\rangle_{\hat n}=\sum_{i=1}^{L+1}\sum_j^{2L}\frac{w_i}{2}\frac{1}{2L+1}\mathcal{I}(f,\iota=\arccos u_i,\phi=\phi_j),
\end{equation}
where $u_i$ and $w_i$ are the $i^\mathrm{th}$ node and corresponding weight of the $n_u$-point Gauss-Legendre rule, where $u_i$ is the $i^\mathrm{th}$ zero of the Legendre Polynomial $P_n(u)$, and 
\begin{equation}
    w_i=\frac{2}{(1-u_i^2)[P'_n(u_i)]}.
\end{equation}
\section{Why decompose $\mathsf{J}_{ts}$ rather than $J_{ts}$?}\label{app:SVD}
Based on the Eckart-Young theorem~\cite{Eckart:1936}, performing the truncated \ac{SVD} algorithm on any arbitrary matrix $M$ yields a matrix $M_R$ with rank $r\leqslant R$ that minimizes the Frobenius error:
\begin{equation}
    M_R={\mathrm{arg}}\underset{\mathrm{rank}(\tilde M)\leqslant R}{{\min}}\sum_{x,k}(M_{xk}-\tilde M_{xk})^2.
\end{equation}
This theorem shows that truncated \ac{SVD} weights every matrix entry equally. Consequently, the choice of matrix determines the target function being minimized.

Our ultimate goal is to make sure that the relative error of the evaluated $\OGW(f)$ is small at \emph{every frequency} and for \emph{any population}. The truncation error enters Eq.~\eqref{eq:summation_stage1} in the following way:
\begin{equation}
    \delta\OGW(f_k)=\frac{4\pi^2f_k^3}{3H_0^2}\delta^2\sum_de^{4\ell_{\det,d}}\sum_s \varpi_s K_{d,s}\delta J_{d+k,s}.
\end{equation}
As a reminder, $\varpi_s$ denotes the Simpson's weights defined in Eq.~\eqref{eq:varpi}.

For a given $d$, we have the error term $\varepsilon$
\begin{equation}
    \varepsilon\equiv \sum_s\varpi_sK_{d,s}\delta J_{t,s},
\end{equation}
where $t\equiv d+k$ is the index of $x$.

By the Cauchy-Schwarz inequality, we have
\begin{equation}
    \left|\sum_s\varpi_sK_{d,s}\delta J_{t,s}\right|\leqslant \sqrt{\sum_s \varpi_sK_{d,s}^2}\sqrt{\sum_s\varpi_s\delta J_{t,s}^2}.
\end{equation}
The first term on the right-hand side only depends on Simpson's weight and population, so what we can control is only the second term. Minimizing the second term therefore bounds the error uniformly for any arbitrary populations.

We then define the weighted sum of $\delta J_{t,s}^2$ over $\lambda_s$ for a given $t$, $E_t^2$, as follows:
\begin{equation}
    E_t^2\equiv \sum_s\varpi_s\delta J^2_{t,s}.
\end{equation}

The next step is the sum over $d$, or equivalently, $t$. At a fixed observation frequency $f_k$, Eq.~\eqref{eq:summation_stage1} collects the contribution of all detector-frame masses. A single frequency bin therefore uses rows of the $J(x;\lambda)$ table ranging from the inspiral of light systems to the merger and ringdown of heavy ones. At $f=100\,\mathrm{Hz}$, for example, a $2.8\,M_\odot$ binary gives $x\simeq 1.4\times 10^{-3}$, while a $200\,M_\odot$ binary gives $x\simeq0.1$, where $J$ is smaller by five orders of magnitude. Its contribution to $\OGW(f_k)$, however, carries the factor $e^{4\ell_{\det}}=M_{\det}^4$, larger by seven orders of magnitude, so the two contributions can be comparable. The magnitude of $J$ in a row thus does not determine the importance of that row: the importance is set by the population, which can make any row relevant for a specific frequency.

To make this explicit, we factor out the scale of each row,
$J_{t,s}=\varsigma_t\,a_{t,s}$, where
$\varsigma_t\equiv\sqrt{\sum_sJ_{t,s}^2/n_q}$ is the root-mean-square of row $t$ over the columns and $a_{t,s}\simeq\mathcal{O}(1)$ is its shape. The contribution of
mass $M_{\det}$ at $f_k$ is then
\begin{equation}
\sum_s \varpi_sK_{d,s}J_{t,s}=\varsigma_t\sum_s \varpi_sK_{d,s}\,a_{t,s},
\end{equation}
i.e., the row scale $\varsigma_t$ times a factor that depends only on the population and two quantities of order unity.

To guarantee the relative error in $\OGW(f)$ is small, the truncation error of row $t$ must therefore be small \emph{relative to}
$\varsigma_t$, uniformly in $t$, which motivates the normalized error
\begin{equation}
(E^{\rm norm}_t)^2\equiv\sum_s \varpi_s\,\frac{\delta J_{t,s}^2}{\varsigma_t^2}.
\label{eq:normalized_Et}
\end{equation}
Without this normalization the truncated SVD would distribute the \emph{absolute}
error evenly over the rows, at a level set by the largest ones (the inspiral,
$\varsigma\sim10^{-42}$). The relative error in the merger and ringdown rows
($\varsigma\sim10^{-64}$) would then be enormous, and $\OGW(f_k)$ would be biased wherever heavy systems contribute.

Consequently, we choose to minimize $\sum_t (E^{\rm norm}_t)^2$, the sum of Eq.~\eqref{eq:normalized_Et} over all rows of the $J(x;\lambda)$ table. This is then exactly the Frobenius error of the matrix $\mathsf J_{ts}\equiv \sqrt{\varpi_s}\,J_{t,s}/\varsigma_t$, so by the Eckart-Young theorem it is minimized by the truncated \ac{SVD} of $\mathsf J$.
\section{Preconditioning before cross-correlation}\label{app:preconditioning}
We use \ac{FFT} to compute the cross-correlation $c_k$ of $\hat K_{d,r}$ and $B_{d+k,r}$.
Then $c_{k,r}=(\hat K\star B)_{k,r}$ is  computed at once for all $k$ indices via \ac{FFT}. 

We highlight that the round-off error of an \ac{FFT}-based correlation is uniform across the entries. For a given mode $r$, its magnitude is therefore determined by the largest entry of $c_k$. We note that $\hat K_{d,r}$ is determined by the redshifted mass distribution, and therefore we cannot do too much about it. However, $B_{d+k,r}\sim J_{d+k,r}\sim x_{d+k}^{-7/3}$ in the inspiral phase. The magnitude of $B$ can cross tens of orders of magnitude for different values of $k$. Consequently, $c_{k,r}$ will also have a large dynamical range. If one simply uses \ac{FFT} to compute $c_{k,r}$, then the relative error of $c_k$ may be extremely large for large $k$'s, i.e., high frequencies. Since we know the trend of $J\sim x^{-7/3}$ at inspiral, we can remove this trend before we do the cross-correlation. We therefore multiply $B$ by a factor of $x^{A}$, where $A=7/3$. Then we have:
\begin{equation}
e^{4\ell_{det}}B_{d+k,r}=(f_kT_\odot)^{-A}\left[e^{(4-A)\ell_{\det,d}}\left(x^A_{d+k}B_{d+k,r}\right)\right],
\end{equation}
and we perform \ac{FFT} on the term in the bracket. It is worth noting that applying this preconditioning ``flattens'' the power of $e^{\ell_{\det,d}}$ from $4$ to $4-A=5/3$, which also effectively decreases the dynamical range of this term.
Eventually, we obtain Eq.~\eqref{eq:summation_mz}.
\section{Using scrambled-Sobel points to perform \ac{QMC}}\label{app:QMC}
Suppose we aim to compute the integral of $f(x)$ over the range $[0,\,1]$:
\begin{equation}
    I\equiv \int_0^1 f(x)\D x.
\end{equation}
One usually leverages \ac{MC} method by randomly scattering $N_\mathrm{samp}$ points $x_i\sim\mathcal{U}[0,1]$ and approximates the integral by
\begin{equation}
    \hat I=\frac{1}{N_\mathrm{samp}}\sum_{i=1}^{N_\mathrm{samp}}f(x_i).
    \label{eq:MC}
\end{equation}
Since these $N_\mathrm{samp} $ points are independent, the variance of $\hat I$ is
\begin{equation}
    \sigma^2_{\hat I}=\frac{1}{N^2_\mathrm
    {samp}}\sum_{i=1}^{N_\mathrm{samp}}\sigma^2_f=\frac{\sigma_f^2}{N_\mathrm{samp}},
\end{equation}
where $\sigma^2_f$ is defined by
\begin{equation}
    \sigma_f^2\equiv \int_0^1f(x)^2\D x-\left(\int_0^1f(x)\D x\right)^2.
\end{equation}
Then as $N_\mathrm{samp}$ increases, the variance of the estimator decrease as $1/N_\mathrm{samp}$, i.e., its root-mean-square error falls as $N_{\rm samp}^{-1/2}$.

\begin{figure*}[!htbp]
    \centering
    \includegraphics[width=0.9\linewidth]{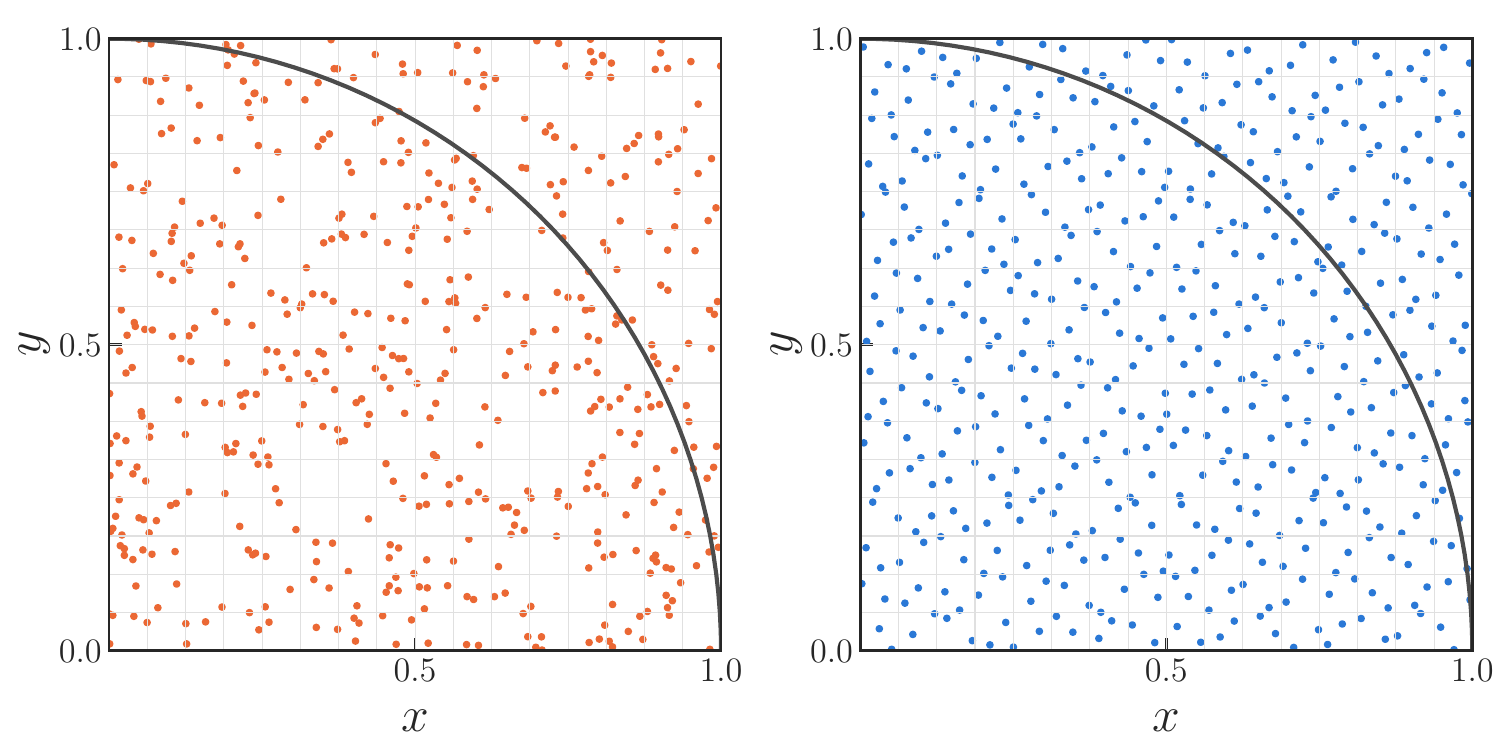}
    \caption{Example of samples used for Monte-Carlo and Quasi Monte-Carlo summations. Left: truly random samples drawn uniformly from $x_i,y_i\sim\mathcal{U}[0,1]^2$. Right: Scrambled Sobel points.}
    \label{fig:MC_QMC_Scatter}
\end{figure*}

\begin{figure*}[!htbp]
    \centering
\includegraphics[width=0.9\linewidth]{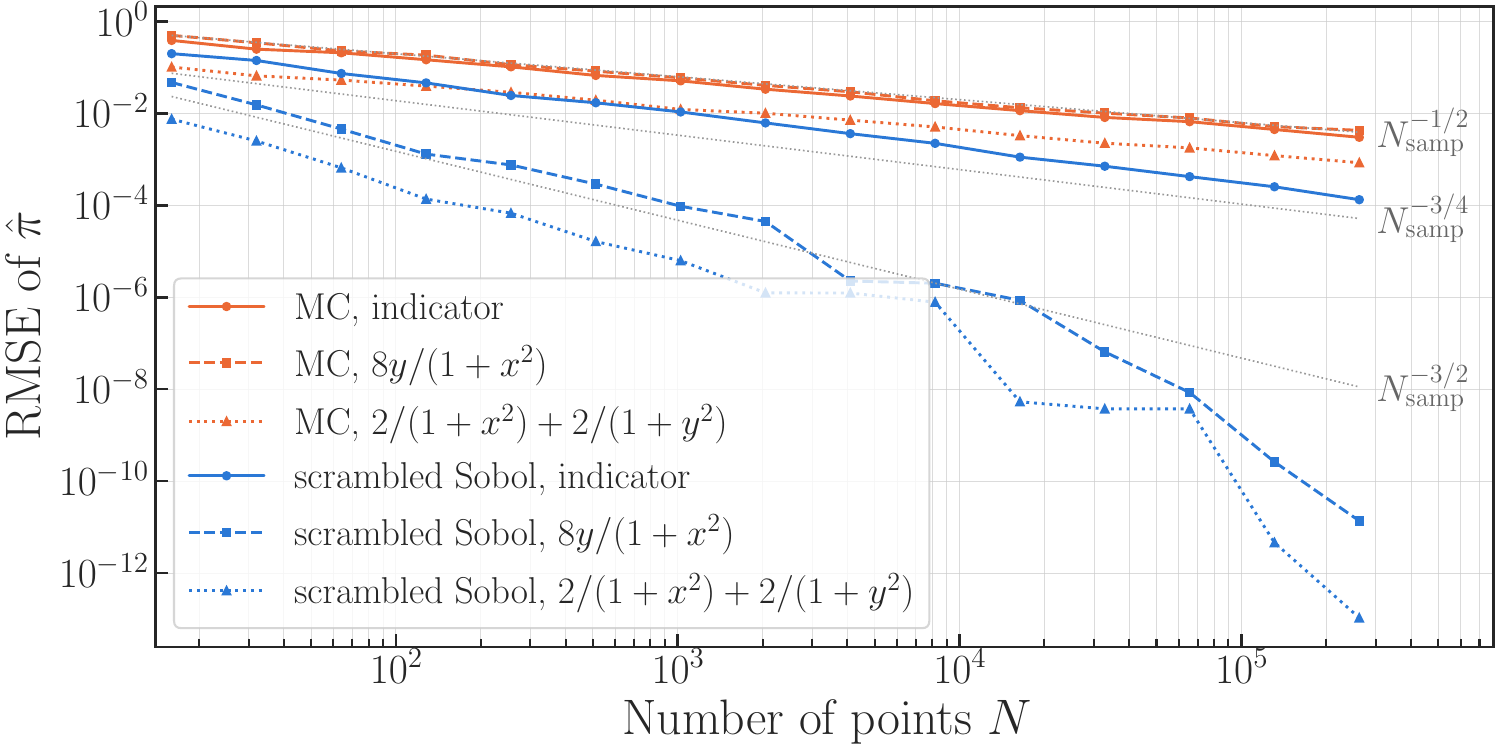}
    \caption{Root-mean-square-error of the $\hat \pi$ estimators for two methods corresponding to different integrand. The convergence speed for the \ac{MC} method is always consistent with $N_{\rm samp}^{-1/2}$ for all three integrands, whereas the convergence speed for the \ac{QMC} method is faster and different for different integrands.}
    \label{fig:MC_QMC_convergence}
\end{figure*}

The \ac{QMC} method also evaluates integrals with Eq.~\eqref{eq:MC}, but replaces the random samples with a deterministic \emph{low discrepancy} sequence, such as the Sobol sequence~\cite{Sobol:1967}, whose points fill the integration domain more evenly than random points. Since the \ac{QMC} samples are not independent, the error of the estimator cannot be estimated from the sample variance, as is done for normal \ac{MC} samples. Random scrambling solves this problem: each scrambled sequence is a random draw that keeps the discrepancy low, and the estimator is unbiased over estimates $\hat I^{(1)},\cdots,\hat I^{(N_{\rm samp})}$. Their mean, $\bar{\hat I}\equiv \langle I^{(i)}\rangle_i$, is the estimate of the integral, and its uncertainty is the standard deviation of $\hat I^{(i)}$ divided by $\sqrt N_{\rm samp}$. We note that for a randomly scrambled sequence and a smooth integrand in $d$ dimensions, the root-mean-square error of the \ac{QMC} estimator falls as $\mathcal{O}((\ln N_{\rm samp})^{(d-1)/2}N_{\rm samp}^{-3/2})$~\cite{Owen1997AoS}.

Here, to show the convergence speed of \ac{QMC}, we present a simple example using \ac{MC} and \ac{QMC} to estimate $\pi$. 

We show how points are drawn for \ac{MC} and \ac{QMC} in Fig.~\ref{fig:MC_QMC_Scatter}, where the left panel shows the plain \ac{MC} method, in which each point is independently drawn from a uniform distribution $x,y\sim\mathcal{U}[0,1]^2$, and the right panel shows the scrambled Sobol points.

To evaluate $\pi$, we consider three integrands, and hence three estimators of $\pi$:

\begin{equation}
\begin{aligned}
&\hat \pi_1=\frac{1}{N_{\rm samp}}\sum_{i=1}^{N_{\rm samp}}\mathbbm{1}(x_i^2+y_i^2\leqslant 1),\\
&\hat \pi_2=\frac{1}{N_{\rm samp}}\sum_{i=1}^{N_{\rm samp}}\frac{8y_i}{1+x^2_i},\\
&\hat \pi_3=\frac{1}{N_{\rm samp}}\sum_{i=1}^{N_{\rm samp}}\left(\frac{2}{1+x_i^2}+\frac{2}{1+y_i^2}\right).
\end{aligned}
\end{equation}

We show the results in Fig.~\ref{fig:MC_QMC_convergence}, where the orange curves correspond to the results using the plain \ac{MC} method, and the blue ones to the \ac{QMC} method. We observe that for the plain \ac{MC} method, the convergence speed is always proportional to $N_{\rm samp}^{-1/2}$, while this is not true for the \ac{QMC} method, where the convergence speed is also dependent on the integrand itself. In this example, we find that for a smooth integrand, the \ac{QMC} error decays faster than $N_{\rm samp}^{-1}$, whereas for the discontinuous indicator integrand it decays as $N_{\rm samp}^{-3/4}$.

In our case, we use \ac{QMC} to perform the integral
\begin{equation}
    I=\int\D\bm \chi_1\D\bm \chi_2~p(\bm \chi_1,\bm \chi_2|\bm\theta_\mathrm{pop})J(x;\lambda,\bm \chi_1,\bm \chi_2),
\end{equation}
where $J(x;\lambda,\bm \chi_1,\bm \chi_2)$ is a smooth function of the waveform, and the population of the spin parameters is also smooth. In addition to the to the smoothness of the integrand, the dimension of this integral is also low. Consequently, adopting \ac{QMC} can accelerate the speed of convergence of the spin-parameter average.

\begin{figure*}[!htbp]
\centering
\includegraphics[width=\textwidth]{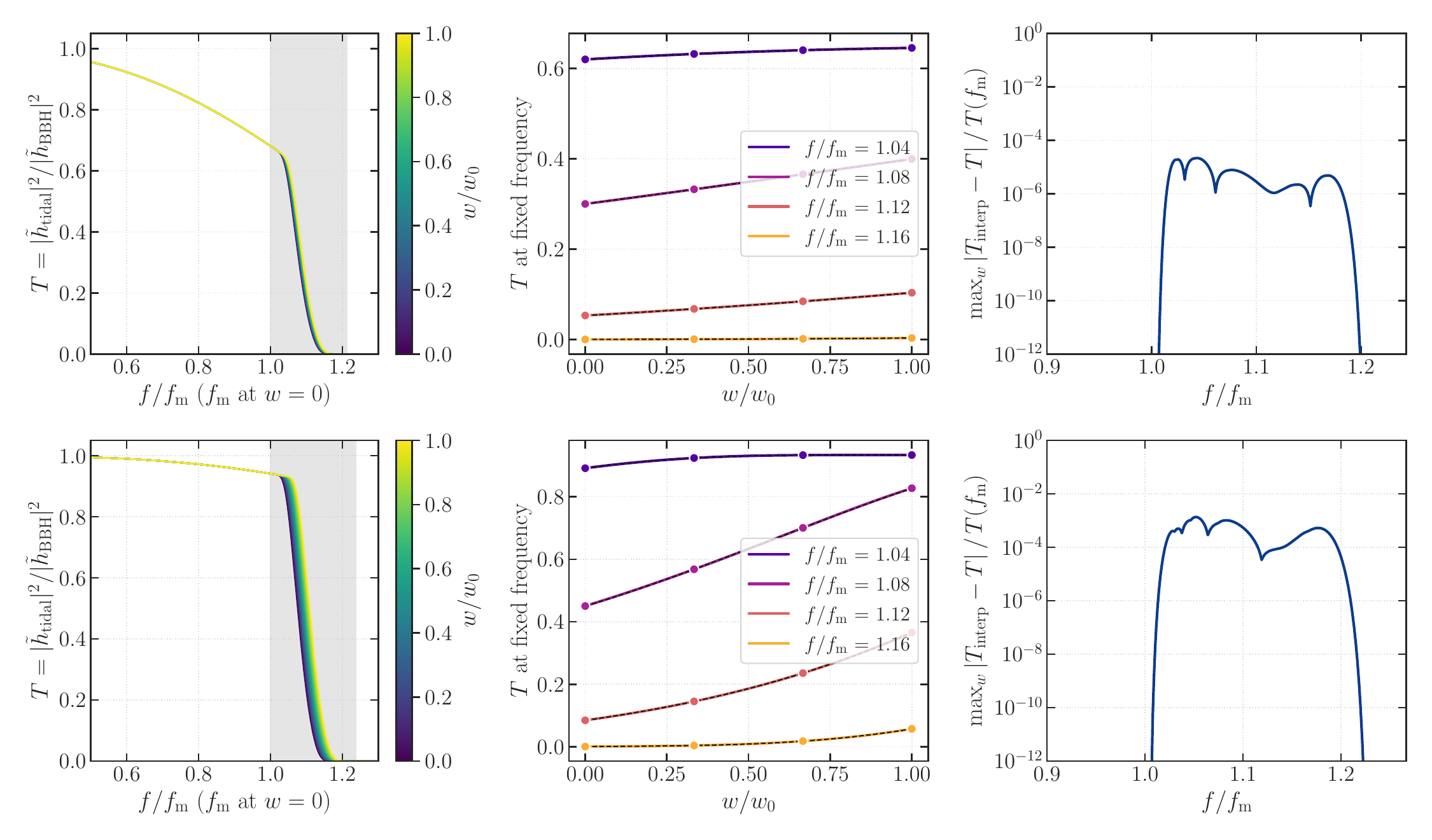}
\caption{Dependence of the \texttt{NRTidalv3} intensity on $w$, at fixed $\kappa_{\rm eff}^T=60$ and $M=2.7\,M_\odot$, for $q=0.7$ (top) and $q=0.3$ (bottom).
Left: Tidal factor $T$ [Eq.~\eqref{eq:tidal_factor}] for $w$ spanning $[0,w_0]$ (color);
grey: the union of the tapers of all $w$. Middle: $T$ as a function of $w$ at fixed frequencies
within the taper: exact (solid), the four nodes (dots), and the cubic interpolation (dashed). 
Right:  Largest error over $w$, in units of
$T(f_{\rm m})$. Frequencies are in units of $f_{\rm m}$ at $w=0$.}
\label{fig:split_taper}
\end{figure*}
\section{Why is $n_w=4$ sufficient?}\label{app:n_w}
In this appendix we justify that $n_w=4$ nodes in $w$ suffice for \texttt{NRTidalv3}. 

\paragraph{$w$ enters only through the merger frequency.}
In our study, only the amplitude $|\tilde h|^2$ matters for the $\OGW(f)$ computation. \texttt{NRTidalv3} models the tidal amplitude correction in the same way as \texttt{NRTidalv2}, and is only dependent on $\kappa_{\rm eff}^T$ (cf. Eqs.~(29) and (40) of Ref.~\cite{Abac:2023ujg}). The effects that depend on the individual tidal deformabilities enter the phase correction only. The individual
deformabilities, and hence $w$, enter the amplitude only through the merger frequency $f_{\rm m}$: above $f_{\rm m}$ the waveform is tapered smoothly to zero, which it reaches at $1.2 f_{\rm m}$ (cf. Eq.~(46) of Ref.~\cite{Abac:2023ujg}). We can therefore write the tidal factor, the ratio of
the power with tides to that of the underlying point-particle model, as
\begin{equation}
T(f;\kappa_{\rm eff}^T,w)\equiv\frac{|\tilde h_{\rm tidal}|^2}{|\tilde h_{\rm BBH}|^2}
=A^2(f;\kappa_{\rm eff}^T)\,\Sigma^2\!\Big(\frac{f}{f_{\rm m}(\kappa_{\rm eff}^T,w)}\Big),
\label{eq:tidal_factor}
\end{equation}
where $A$ is the amplitude correction, and $\Sigma$ a Planck taper that equals one below
$f_{\rm m}$ and falls smoothly to zero at $1.2f_{\rm m}$.

\paragraph{$f_{\rm m}$ depends weakly and smoothly on $w$.}
The merger frequency of \texttt{NRTidalv3} is a fit in terms of $\kappa_{2}^T$
[Eqs.~(41)--(44) of Ref.~\cite{Abac:2023ujg}],
\begin{equation}
\begin{aligned}
\kappa_{2}^T&=3X_AX_B[X_A^3\Lambda_A+X_B^3\Lambda_B]\\
&\equiv\kappa_{\rm eff}^T\,[\rho_1w+\rho_2(1-w)],
\end{aligned}
\end{equation}
where 
\begin{equation}
    X_{A/B}\equiv\frac{m_{A/B}}{M},\quad \rho_1=\frac{13q}{1+12q},\quad \rho_2=\frac{13}{12+q}.
\end{equation}
Therefore, $\kappa_2^T$ is linear in $w$ at fixed $\kappa_{\rm eff}^T$ and $q$. Since the coefficients $\rho_{1,2}$ are close
to unity ($\rho_1=0.968$, $\rho_2=1.024$ for $q=0.7$), $\kappa_{2}^T$ varies by only a few
percent over the physical range $0\leqslant w\leqslant w_0$. As a result, $f_{\rm m}$ moves by $1.1\%$ for $q=0.7$ and by at most $3.1\%$ for $q=0.3$, when $M=2.7M_\odot$ and $\kappa_{\rm eff}^T=60$ (i.e., $\tilde\Lambda=320$).

\paragraph{The shift is small compared with the width of the taper.}
The taper spans $f_{\rm m}\leqslant f\leqslant 1.2f_{\rm m}$, most of the drop occurring between $1.05f_{\rm m}$ and $1.15f_{\rm m}$. 
Since $w$ moves $f_m$ by only a small fraction of this width, $T$ at a given frequency is a smooth, slowly varying function of $w$, and a  cubic polynomial in $w$ then reproduces it
to high accuracy.

In Fig.~\ref{fig:split_taper} we show these properties for one binary ($M=2.7\,M_\odot$,
$\kappa_{\rm eff}^T=60$) with a typical mass ratio, $q=0.7$ (top), and with the smallest mass
ratio of the grid, $q=0.3$ (bottom), which is the most demanding case. Frequencies are in units
of the merger frequency for the $w=0$ case.
The left panel shows $T$ for values of $w$ spanning the physical range. Below the merger the curves
coincide exactly, as stated by Eq.~\eqref{eq:tidal_factor}: $T$ decreases slowly due to the
amplitude correction, independently of $w$. The curves separate only within the taper (gray;
the union of the tapers of all $w$), where varying $w$ shifts the taper horizontally, by $1.1\%$
of $f_{\rm m}$ for $q=0.7$ and $3.1\%$ for $q=0.3$.
The middle panel cuts the left panel vertically: at fixed frequencies within the taper, it shows $T$ as a
function of $w$ (solid lines), together with the values at the four nodes of the table (dots) and the cubic interpolation through them (dashed lines). $T$ is a smooth function of $w/w_0$; it is flat near the start of the taper ($f=1.04f_{\rm m}$), steepest where the taper falls fastest, and vanishes at its bottom. The interpolation cannot be distinguished from the exact curves, even for $q=0.3$, where the larger shift makes $T(w)$ visibly curved.
The right panel quantifies the error as the largest deviation over $w$ in units of $T$ at the merger. It vanishes identically below the merger and is confined to the taper. The cubic interpolation on four nodes results in a relative error of $2\times10^{-5}$ and $1\times10^{-3}$ for the $q=0.7$ and $q=0.3$ cases, respectively.

\paragraph{Error in $\Omega_{\rm GW}$.}
The errors of the right panel refer to a single binary, a single frequency and the worst $w$. The error in $\Omega_{\rm GW}$ is much smaller: at a given detector-frame frequency, only the binaries whose taper falls there contribute to it, and the interpolation error changes sign across the taper (the dips in the right panel), so that it largely cancels in the sum over masses and redshifts. Four nodes
thus make this error negligible compared with the other numerical errors of the method
($\sim10^{-4}$, Sec.~\ref{sec:convergence}).
\section{Supported models and default parameters}\label{app:supported}
In this appendix we present tables summarizing the waveform models, population models, and some default parameters adopted in \texttt{StochAXtic}. Table~\ref{tab:waveforms} summarizes the main waveform models currently implemented in \texttt{StochAXtic}. Table~\ref{tab:populations} and Table~\ref{tab:populations2} summarize the mass, redshift, spin, and \ac{EOS} models supported by \texttt{StochAXtic}. In Table~\ref{tab:grids} and Table~\ref{tab:defaults} we summarize the default numerical settings of \texttt{StochAXtic}.
\providecommand{\code}[1]{\texttt{#1}}
\providecommand{\tbc}{\textsuperscript{$\dagger$}}
\providecommand{\arxiv}[1]{arXiv:#1}
\providecommand{\tfrac}{\frac}
\providecommand{\toprule}{\hline}
\providecommand{\midrule}{\hline}
\providecommand{\bottomrule}{\hline}

\begin{table*}
\centering
\footnotesize
\renewcommand{\arraystretch}{1.2}
\begin{tabular}{ccccccc}
\hline\hline
Approximant & Source & Spins & Tides & Modes & $\langle \mathcal{I}\rangle_{\hat n}$ & Reference \\
\hline
\code{IMRPhenomD}              & all & aligned    & ---                                & $(2,2)$ & face-on & Refs.~\cite{Khan:2015jqa,Husa:2015iqa} \\
\hline
\code{IMRPhenomXAS}            & all & aligned    & ---                                & $(2,2)$ & face-on & Ref.~\cite{Pratten:2020fqn} \\
\hline
\code{IMRPhenomXHM}            & all & aligned    & ---                                & $(2,2)$, $(2,1)$, $(3,3)$, $(3,2)$, $(4,4)$ & $(5,9)$ & Ref.~\cite{Garcia-Quiros:2020qpx} \\
\hline
\code{IMRPhenomXP}             & all & precessing & ---                                & $(2,2)$ & $(3,5)$ & Ref.~\cite{Pratten:2020ceb} \\
\hline
\code{IMRPhenomXPHM}           & all & precessing & ---                                & $(2,2)$, $(2,1)$, $(3,3)$, $(3,2)$, $(4,4)$ & modes & Ref.~\cite{Pratten:2020ceb} \\
\hline
\code{IMRPhenomPv3}            & all & precessing & ---                                   & $(2,2)$ & co-prec. & Ref.~\cite{Khan:2018fmp} \\
\hline
\code{IMRPhenomPv3HM}          & all & precessing & ---                                   & $(2,2)$, $(2,1)$, $(3,3)$, $(3,2)$, $(4,4)$, $(4,3)$ & co-prec. & Ref.~\cite{Khan:2019kot} \\
\hline
\code{SEOBNRv5\_ROM}           & all & aligned    & ---                                   & $(2,2)$ & face-on & Ref.~\cite{Pompili:2023tna} \\
\hline
\code{SEOBNRv5HM\_ROM}         & all & aligned    & ---                                   & $(2,2)$, $(2,1)$, $(3,3)$, $(3,2)$ $(4,4)$, $(4,3)$, $(5,5)$& modes & Ref.~\cite{Pompili:2023tna} \\
\hline
\code{IMRPhenomXAS\_NRTidalv2} & BNS            & aligned    & $\kappa_{\rm eff}^T$               & $(2,2)$ & face-on & Ref.~\cite{Dietrich:2019kaq} \\
\hline
\code{IMRPhenomXP\_NRTidalv2}  & BNS            & precessing & $\kappa_{\rm eff}^T$               & $(2,2)$ & $(3,5)$ & Refs.~\cite{Dietrich:2019kaq,Pratten:2020ceb} \\
\hline
\code{IMRPhenomXAS\_NRTidalv3} & BNS            & aligned    & $\kappa_{\rm eff}^T$, $\kappa_2^T$ & $(2,2)$ & face-on & Ref.~\cite{Abac:2023ujg} \\
\hline
\code{IMRPhenomXP\_NRTidalv3}  & BNS            & precessing & $\kappa_{\rm eff}^T$, $\kappa_2^T$ & $(2,2)$ & $(3,5)$ & Refs.~\cite{Abac:2023ujg,Pratten:2020ceb} \\
\hline
\code{SEOBNRv5\_ROM\_NRTidalv3} & BNS           & aligned    & $\kappa_{\rm eff}^T$, $\kappa_2^T$    & $(2,2)$ & face-on & Refs.~\cite{Pompili:2023tna,Abac:2023ujg} \\
\hline
\code{IMRPhenomNSBH}           & NSBH           & aligned    & $\kappa_{\rm eff}^T$               & $(2,2)$ & face-on & Ref.~\cite{Thompson:2020nei} \\
\hline\hline
\end{tabular}
\caption{Waveform approximants (LALSuite). For the precessing models, the modes are those in the
co-precessing frame, which follows the precession of the orbital plane. The orientation average
$\langle \mathcal I\rangle_{\hat n}$ is computed exactly: ``face-on'' uses $\langle \mathcal I\rangle_{\hat n} = \tfrac{2}{5} \mathcal{I}(\iota=0)$ (see App.~\ref{app:faceon});
$(n_u, n_\phi)$ is a Gauss product rule in $(\cos\iota, \phi)$, exact for the modes the model contains (see App.~\ref{app:general});
``modes'' sums the mode intensities analytically (see App.~\ref{app:modes});
``co-prec." refers to the  summation over modes in the co-precessing frame discussed in App.~\ref{app:modes}. Other LAL frequency-domain approximants
(e.g., \code{IMRPhenomPv2}) use the default $(n_u=5,n_\phi=9)$ rule, exact for $\ell\leqslant 4$.}
\label{tab:waveforms}
\end{table*}

\providecommand{\cell}[1]{\begin{tabular}{@{}c@{}}#1\end{tabular}}

\begin{table*}
\centering
\footnotesize
\renewcommand{\arraystretch}{1.2}
\begin{tabular}{ccccc}
\hline\hline
Name & Source & Form & Defaults & Reference \\
\hline
\code{power\_law\_plus\_peak} & BBH
 & \cell{$p(m_1)\propto[(1-\lambda)\,m_1^{-\alpha}+\lambda\,\mathcal N(\mu_m,\sigma_m)]\times$taper, $m_1\leqslant m_{\max}$\\ $p(m_2|m_1)\propto m_2^{\beta}\times$ Planck taper}
 & GWTC-3.0 fit & Refs.~\cite{Talbot:2018cva,LIGOScientific:2020kqk,KAGRA:2021duu} \\
\hline
\code{bpl\_double\_gaussian} & BBH
 & \cell{broken power law $+$ two Gaussian peaks in $m_1$\\ $p(m_2|m_1)\propto m_2^{\beta}$; separate Planck tapers for $m_1$, $m_2$}
 & GWTC-5.0 fit & Refs.~\cite{LIGOScientific:2025pvj,LIGOScientific:2026ctl} \\
\hline
\code{uniform} & BNS & uniform on $m_{\min}\leqslant m_2\leqslant m_1\leqslant m_{\max}$ & $[1,2]\,M_\odot$ & Refs.~\cite{KAGRA:2021duu,Landry:2021hvl} \\
\hline
\code{nsbh} & NSBH & $p(m_1)\propto 1/m_1$ (BH), $m_2$ uniform (NS) & \cell{BH $[3,50]\,M_\odot$\\ NS $[1,2]\,M_\odot$} & Refs.~\cite{KAGRA:2021duu,Ebersold:2025izh} \\
\hline
\code{nsbh\_powerlaw\_gaussian} & NSBH & \cell{$p(m_1)\propto m_1^{-\alpha_{\rm BH}}$ on $[m_{\rm BH,min},m_{\rm BH,max}]$\\ $p(q|m_1)$ truncated Gaussian}
 & \cell{$\alpha_{\rm BH}=2.51$\\ $m_{\rm BH}\in[3.31,13]\,M_\odot$} & Ref.~\cite{Biscoveanu:2022iue} \\
\hline\hline
\end{tabular}
\caption{Population models: mass distributions (continued in Table~\ref{tab:populations2}). All models
have JAX implementations in \texttt{StochAXtic} (differentiable in every hyperparameter). The Planck taper has a smooth low-mass turn-on of width $\delta_m$.}
\label{tab:populations}
\end{table*}

\begin{table*}
\centering
\footnotesize
\renewcommand{\arraystretch}{1.2}
\begin{tabular}{ccccc}
\hline\hline
Name & Source & Form & Defaults & Reference \\
\hline
\multicolumn{5}{c}{\bf{\emph{Merger-rate models} $R(z)$}} \\
\hline
\code{madau\_dickinson} & all & \cell{$R_0\,\psi(z)/\psi(0)$\\ $\psi=(1+z)^{\alpha}/[1+((1+z)/(1+z_p))^{\beta}]$}
 & \cell{$R_0\simeq18\,{\rm Gpc^{-3}\,yr^{-1}}$\\ $\alpha=2.85$, $z_p=2.27$, $\beta=6.99$} & \cell{Ref.~\cite{Madau:2014bja} (form)\\ Ref.~\cite{LIGOScientific:2026ctl} (defaults)} \\
\hline
\code{madau\_dickinson\_timedelay} & BNS, NSBH & \cell{star-formation rate convolved with\\ $P(t_d)\propto 1/t_d$, $t_d\in[20\,{\rm Myr}, t_H]$}
 & \cell{$R_0=23.4\,{\rm Gpc^{-3}\,yr^{-1}}$\\ $\alpha=2.6$, $z_p=2.2$, $\beta=6.2$} & \cell{Ref.~\cite{Madau:2014bja} (SFR)\\ Ref.~\cite{KAGRA:2021kbb} (delays)} \\
\hline
\code{power\_law} & all & $R_0(1+z)^{\lambda}$ & $\lambda=2$ & Ref.~\cite{Fishbach:2017zga,Fishbach:2018edt} \\
\hline
\code{constant} & all & $R_0$ & --- & --- \\
\hline
\multicolumn{5}{c}{\bf{\emph{Spin models} $p(a_1,a_2,\cos t_1,\cos t_2)$, azimuths uniform}} \\
\hline
\code{gwtc3\_default} & BBH & \cell{$a_i/a_{\max}\sim{\rm Beta}(\alpha_\chi,\beta_\chi)$\\ $\zeta\,\mathcal N(1,\sigma_t)\,\mathcal N(1,\sigma_t)+(1-\zeta)/4$}
 & \cell{$\alpha_\chi=1.64$, $\beta_\chi=4.15$\\ $\zeta=0.66$, $\sigma_t=1.54$} & Refs.~\cite{Wysocki:2018mpo,KAGRA:2021duu} \\
\hline
\code{gwtc5\_default} & BBH & \cell{$a_i\sim\mathcal N(\mu_\chi,\sigma_\chi)$ on $[0,a_{\max}]$\\ $\zeta\,\mathcal N(\mu_t,\sigma_t)\,\mathcal N(\mu_t,\sigma_t)+(1-\zeta)/4$}
 & \cell{$\mu_\chi=0.063$, $\sigma_\chi=0.37$\\ $\zeta=0.67$, $\mu_t=0.32$, $\sigma_t=1.11$} & Refs.~\cite{LIGOScientific:2026ctl,Vitale:2022dpa} \\
\hline
\code{uniform\_isotropic} & BNS & \cell{$a_i\sim\mathcal U[0,a_{\max}]$\\ $\cos t_i\sim\mathcal U[-1,1]$ (isotropic)}
 & \cell{no hyperparameters\\ $a_{\max}=0.05$ (table bound)} & \cell{LVK low-spin prior\\ Ref.~\cite{LIGOScientific:2017vwq}} \\
\hline
\multicolumn{5}{c}{\emph{Tide model} $\Lambda(m)$} \\
\hline
\code{eos\_lnpoly} & BNS, NSBH & \cell{$\ln\Lambda=\ln\Lambda_{1.4}+b_1\ln(m/1.4)$\\ $+\,b_2\ln^2(m/1.4)$; $\Lambda=0$ above $m_{\rm NS,max}$}
 & \cell{$\Lambda_{1.4}=400$, $b_1=-6$\\ $b_2=-3$, $m_{\rm NS,max}=2.3\,M_\odot$} & \cell{this work\\ cf.\ Ref.~\cite{DelPozzo:2013ala}} \\
\hline\hline
\end{tabular}
\caption{Merger-rate, spin and tide models (continuation of Table~\ref{tab:populations}).}
\label{tab:populations2}
\end{table*}

\providecommand{\cell}[1]{\begin{tabular}{@{}c@{}}#1\end{tabular}}

\begin{table*}
\centering
\footnotesize
\renewcommand{\arraystretch}{1.2}
\begin{tabular}{cccc}
\hline\hline
Parameter & Meaning & Default & Convergence test \\
\hline
$\delta$ (\code{delta}) & spacing of the $\ln M$, $\ln(1+z)$ and $\ln f$ grids & $0.005$ & $0.0025$ \\
\hline
\code{table\_step} & spacing of the $\ln x$ grid of the $J$ table, in units of $\delta$ & $2$ & $1$ \\
\hline
$n_q$ (\code{n\_q}) & Simpson nodes in $\ln q\in[\ln q_{\min},0]$ & \cell{$121$ (BBH)\\ $41$ (BNS, NSBH)} & $2n_q-1$ \\
\hline
$n_\chi$ (\code{n\_spin}) & scrambled Sobol spin samples per $q$ node & $128$ & \cell{$256$\\ new scrambling} \\
\hline
$g(a)$ & \cell{proposal of the spin magnitudes\\ (tilts and azimuths uniform)}
 & \cell{$0.3\,\mathcal U[0,1]+0.7\,\mathcal N_{[0,1]}(0.15,0.45)$ (BBH)\\ $\mathcal U[0,a_{\max}]$ (BNS, NSBH)} & --- \\
\hline
$R$ (\code{rank}) & SVD rank: $\sigma_{R+1}/\sigma_1<10^{-4}$, $R\leqslant96$ & set by \code{rank\_tol} $=10^{-4}$ & $2R$ \\
\hline
$n_\kappa$ (\code{n\_kappa}) & \cell{nodes in $\kappa_{\rm eff}^T$: $0$ and $n_\kappa-1$ log-spaced\\ cubic in $\ln\kappa_{\rm eff}^T$}
 & \cell{$24$ on $[0.1,3000]$ (NRTidal)\\ $48$ on $[0.01,300]$ (\code{IMRPhenomNSBH})} & \cell{$48$\\ $96$} \\
\hline
$n_w$ (\code{n\_split}) & nodes in $w$ (NRTidalv3) on $[0,w_0(q)]$; cubic Lagrange & $4$ & $7$ \\
\hline
$n_{\rm sub}$ (\code{n\_sub}) & sub-samples of the mass density per $\ln M$ cell & \cell{$4$\\ $1$ for BP2P (smooth edges)} & $2n_{\rm sub}$ \\
\hline
$[z_{\min},z_{\max}]$ & redshift range & $[0.2,10]$ & --- \\
\hline
cosmology & --- & Planck18 & --- \\
\hline\hline
\end{tabular}
\caption{Default numerical settings of \texttt{StochAXtic}. The last column gives the value used to estimate
the corresponding error in the convergence test (Sec.~\ref{sec:convergence}), where each parameter is changed in turn,
with all others at their defaults.}
\label{tab:defaults}
\end{table*}

\begin{table*}
\centering
\footnotesize
\renewcommand{\arraystretch}{1.2}
\begin{tabular}{cccccccc}
\hline\hline
Source & $M$ range [$M_\odot$] & $q_{\min}$ & $n_q$ & $m_{\rm ref}$ [$M_\odot$] & $x_{\rm ref}$ & $a_{\max}$ & Spin proposal \\
\hline
BBH  & $[1.5, 700]$ & $0.01$ & $121$ & $20$  & $0.005$ & $1$ & defensive \\
\hline
BNS  & $[1.8, 6]$   & $0.3$  & $41$  & $2.8$ & $3\times10^{-4}$ & $0.05$ & uniform \\
\hline
NSBH & $[3.5, 16]$  & $0.06$ & $41$  & $2.8$ & $10^{-3}$ & $1$ (BH), $0.05$ (NS) & uniform \\
\hline\hline
\end{tabular}
\caption{Grid of the $J$ table for each source class. Here $m_{\rm ref}$ is the total mass at which the table is
computed (scale invariance, Sec.~\ref{sec:convergence}); the grid must contain the support of the population for every
hyperparameter of the prior.}
\label{tab:grids}
\end{table*}

\clearpage
\bibliography{references}

\end{document}